\documentclass[lettersize,journal]{IEEEtran}
\usepackage{amsmath,amssymb,amsfonts,bbm}

\usepackage{amsthm}

\usepackage{algorithmic}
\usepackage{array}
\usepackage[caption=false,font=normalsize,labelfont=sf,textfont=sf]{subfig}
\usepackage{textcomp}
\usepackage{stfloats}
\usepackage{url}
\usepackage{verbatim}
\usepackage{graphicx}
\def\BibTeX{{\rm B\kern-.05em{\sc i\kern-.025em b}\kern-.08em
    T\kern-.1667em\lower.7ex\hbox{E}\kern-.125emX}}
\usepackage{balance}

\usepackage{cleveref}
\crefname{claim}{Claim}{claims}
\crefname{theorem}{Theorem}{theorems}
\crefname{definition}{Definition}{definitions}
\crefname{lemma}{Lemma}{lemmas}
\crefname{fact}{Fact}{facts}
\crefname{corollaryc}{Corollary}{corollaries}
\crefname{remarkc}{Remark}{remarks}
\crefname{remarkt}{Remark}{remarks}
\crefname{remarkd}{Remark}{remarks}
\crefname{remarkl}{Remark}{remarks}
\crefname{figure}{Figure}{fig}

\newcommand{\indep}{\rotatebox[origin=c]{90}{$\models$}}
\newcommand{\sgn}[1]{\operatorname*{sgn}\left( {#1} \right)}
\newcommand{\tr}[1]{\operatorname*{tr}\left( {#1} \right)}
\newcommand{\bs}[1]{\boldsymbol{#1}}
\newcommand{\prob}[1]{\mathbb{P}\left\lbrace {#1} \right\rbrace}
\newcommand{\expc}[3]{\mathbb{E}_{#1}^{#2} \left[ {#3} \right]}
\newcommand{\var}[3]{\mathbb{V}ar_{#1}^{#2} \left[ {#3} \right]}
\newcommand{\dive}[3]{D_{#1} \left( {#2} || {#3} \right)}
\newcommand{\cov}[3]{\text{cov}_{#1}^{#2} \left[ {#3} \right]}
\newcommand{\DR}[2]{%
  \mathrel{\mathop\gtrless\limits^{#1}_{#2}}%
}
\newcommand{\indicator}[1]{\mathbbm{1}_{\left\lbrace {#1} \right\rbrace}}

\newtheorem{definition}{Definition}
\newtheorem{theorem}{Theorem}
\newtheorem{claim}{Claim}
\newtheorem{fact}{Fact}
\newtheorem{lemma}{Lemma}
\newtheorem{corollaryc}{Corollary}[claim]
\newtheorem{remarkl}{Remark}[lemma]   
\newtheorem{remarkc}{Remark}[claim]   

\usepackage{color}
\newcommand{\revised}[1]{\textcolor{black}{#1}}

\allowdisplaybreaks

\begin{document}
\title{Covertly Controlling a Linear System}
\author{Barak Amihood and Asaf Cohen \\ The School of Electrical and Computer Engineering, Ben-Gurion University of the Negev \\  
\thanks{
Parts of this work will be presented at the IEEE Information Theory workshop, ITW 2022.}}

\markboth{Submitted to IEEE Transactions on Information Forensics and Security}%
{How to Use the IEEEtran \LaTeX \ Templates}

\maketitle

\begin{abstract}
Consider the problem of covertly controlling a linear system. In this problem, Alice desires to control (stabilize or change the \revised{behavior} of) a linear system, while keeping an observer, Willie, unable to decide if the system is indeed being controlled or not. 

\revised{We formally define the problem, under a model where Willie can only observe the system's output.}
Focusing on AR(1) systems, we show that when Willie observes the system's output through a clean channel, an inherently unstable linear system can not be covertly stabilized. However, an inherently stable linear system can be covertly controlled, in the sense of covertly changing its parameter 
\revised{or resetting its memory}. Moreover, we give \revised{positive} and \revised{negative} results for two important controllers: a minimal-information controller, where Alice is allowed to use only $1$ bit per sample, and a maximal-information controller, where Alice is allowed to view the real-valued output. Unlike covert communication, where the trade-off is between rate and covertness, the results reveal an interesting \emph{three--fold} trade--off in covert control: the amount of information used by the controller, control performance and covertness. 

\end{abstract}


\maketitle

\section{Introduction}
\label{sec:introduction}
The main objective in control theory is to develop algorithms that govern or control systems. Usually, the purpose of a selected algorithm is to drive the system to a desired state (or keeping it at a certain range of states) while adhering to some constraints such as rate of information, delay, power and overshoot. Essentially, ensuring a level of control (under some formal definition) subject to some (formally defined) constraint. Traditionally, the controller monitors the system's process and compares it with a reference. The difference between the actual and desired value of the process (the error signal) is analysed and processed in order to generate a control action, which in tern is applied as a feedback to the system, bringing the controlled process to its desired state.

Numerous systems require external control in order to operate smoothly. For example, various sensors in our power, gas or water networks monitor the flow and control it. We use various signaling methods to control cameras in homeland security applications or medical devices. In some cases, the controlling signal is manual, while in others it is an automatic signal inserted by a specific part of the application in charge of control. In this paper, however, we extend the setup to cases where achieving control over the system is not the only goal, and consider the control problem while adding an additional constraint of covertness: either staying undetected by an observer (taking the viewpoint of an illegitimate controller), or being able to detect any control operation (taking the viewpoint of a legitimate system owner). This additional covertness constraint clearly seems reasonable in applications such as security or surveillance systems, but, in fact, recent attacks on medical devices \cite{mahler2018know} and civil infrastructures \cite{doi:10.1177/0361198118756885} stress out the need to account for covert control in a far wider spectrum of applications. As a result, it is natural to ask: Can one covertly control a linear system? If so, in what sense? When is it possible to identify with high probability any attempt to covertly control a system? Do answers to the above question depend on the type of controller used, in terms of complexity or information gathered from the system?

While the problem we define herein is related to covert communication, which was studied extensively in the information theory community \cite{bash_limits,Reliable_Deniable,Achieving_Undetectable_Communication,resolvability_perspective,First_and_Second_Order_Asymptotics_in_Covert_Communication,9035417}, and to security in cyber--physical system \cite{mo_secure_2009,smith_decoupled_2011, venkitasubramaniam_information-theoretic_2015,mo_false_nodate,bai_security_2015,zhang_stealthy_2016,weerakkody_information_2016,bai_data-injection_2017,kung_performance_2017,chen_optimal_2018,an_data-driven_2018,fang_stealthy_2020,barboni_detection_2020,mikhaylenko_stealthy_2021}, key differences immediately arise. First, in covert communication, Alice desires to covertly send a message to Bob. Hence, fixing a covertness constraint, Alice's success is measured in rate -- the number of bits per channel use she is able to send covertly. Herein, Alice's objective is \emph{control}, which we measure in the ability to stabilize the system, or change its parameters. In the cyber--physical setup, on the other hand, most works focus on a stable system and fixed parameters, and the goal of an attacker is to degrade the system's performance (e.g., estimation error) without being detected. The trade--off is hence between stealthiness and the ability to degrade performance. A complete literature survey is given in \Cref{subsection:Scientific_overview} below. 

In the covert control setup we define herein, \emph{we identify an additional dimension, which does not exist in the two--dimensional setups above}. In covert control, the amount of information Alice has to extract from the system in order to carry out her objective also plays a key role. In other words, there is a non-trivial interplay between \emph{three measures: information, control and covertness}, adding depth and a multitude of open problems. 
\subsection{Main Contribution}

\begin{figure}[h]
\centering
\includegraphics[width=0.85\linewidth]{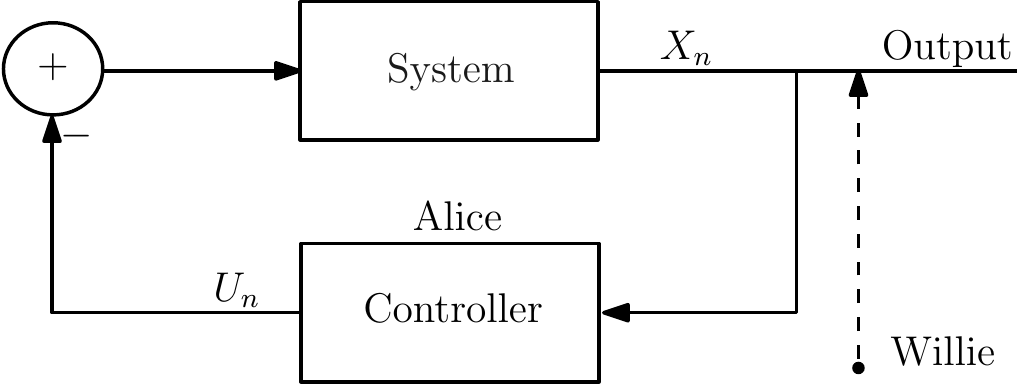}
\caption{A basic Covert Control model. 
Alice observes the system's output, and wishes to control the system, either using a desired reference signal or without it. She can decide how frequently and how accurately to sample the output on the one hand, and how frequent and how strong will be her control signal be, on the other. 
The control signal is an input to the system, but it is also observed by Willie either 
implicitly when observing the output of the system. 
Willie's observations can be either clean, or viewed via a noisy channel. Alice's goal is to control the system without being noticed. Willie's goal is to detect if Alice is indeed controlling the system.}
\label{fig:Covert_Control}
\end{figure}

In this paper, we focus on a simple linear system, schematically depicted in \Cref{fig:Covert_Control}. Alice, observing the system's output $X_n$, wishes to control the system's behaviour though a control signal $U_n$. We assume the system without control follows a first order Auto Regressive model (AR(1)), hence,
\begin{align}\label{eq:simple_linear_model}
\small
X_{n+1} = a X_{n} + Z_{n} - U_{n},
\normalsize
\end{align}
and focus on \revised{three} types of control objectives: (i) Stabilizing an otherwise unstable system. (ii) Changing the parameter, $a$, of a stable system. \revised{(iii) resetting the system's memory}. However, Alice wishes to perform her control action without being detected by Willie. 
\revised{In terms of Willie's observations, Willie can only observe the system's output, and his observations are through a clean channel.}

We formally define covertness and detection in these scenarios. We then show that an unstable AR(1) cannot be covertly controlled, in the sense of maintaining a finite $\gamma$-moment without being detected. 
\revised{
In addition, in the limiting--case of a \emph{minimum--information} controller, in which Alice retrieves only one bit per sample, using only the sign of $X_{n}$ to keep the system bounded, we show that Alice, in fact, cannot covertly control even a stable AR(1). 
On the other hand, we then turn to the opposite limiting case, a \emph{maximum--information} controller, in which Alice is allowed to view the real--valued signal. 
In this case, we consider a Gaussian stable AR(1) system and two different control objectives: parameter change and memory reset. We show that a Gaussian stable AR(1) system can be covertly manipulated by Alice, in the sense of changing its gain without being detected by Willie, and that Alice can manipulate a system, by "resetting" it to its stationary distribution at a time instance of her choice, without being detected by Willie.
In addition, we give a negative result for the latter objective.
}

The results above, to the best of our knowledge, are the first to characterize cases where one can or cannot covertly control a linear system.
Although the results are pinpointed to the two extremes of information usage, 
they 
\revised{shed light on an important three--fold trade--off, yet to be fully characterized, between information, covertness and control, via different operating points.}
\subsection{Related Work}
\label{subsection:Scientific_overview}
\subsubsection{Covert Communication}
Communicating covertly has been a long standing problem. In this scenario, two parties, Alice and Bob, wish to communicate, while preventing a third party, Willie, from detecting the mere presence
of communication. While studied in the steganography and spread spectrum communication literature \cite{Steganography, Spread, bash_hiding}, the first information theoretic investigation was done in \cite{bash_limits, Square_Root_Law}. This seminal work considered Additive White Gaussian Noise channels (whose variances are known to all parties), and computed the highest achievable rate between Alice and Bob, while ensuring Willie's sum of false alarm and missed detection probabilities ($\alpha + \beta$) is arbitrarily close to one. 
This, however, resulted in a transmission rate which is asymptotically negligible, e.g., $O(\sqrt{n})$ bits for $n$ channel uses. In fact, this ``square root law" for covert communication holds more generally, e.g., for Binary Symmetric Channels \cite{Reliable_Deniable} and via channel resolvability \cite{resolvability_perspective}. These works strengthened the insight that forcing $\alpha + \beta$ arbitrarily close to $1$, and assuming Willie uses optimal detection strategies, results in a vanishing rate. In order to achieve a strictly positive covert communication rate, Alice and Bob need some advantage over Willie
\cite{Not_Know_Noise, Not_Know_When, Reliable_deniable_communication_with_channel_uncertainty, Achieving_Undetectable_Communication, Achieving_positive_rate_with_undetectable_communication_over_AWGN_and_Rayleigh_channels, Achieving_positive_rate_with_undetectable_communication_Over_MIMO_rayleigh_channels}, or, alternatively, intelligently use the fact that Willie cannot use an optimal detector in practice, or some key problem parameters are beyond his reach.
For example, \cite{Not_Know_Noise} showed that if Willie does not know his exact noise statistics, Alice can covertly transmit $O(n)$ bits over $T$ transmission slots, one of which she can utilize, where each slot duration can encompass $n$ bit codewords. 
The authors in \cite{Not_Know_When} showed that if Alice and Bob secretly pre-arrange on which out of $T(n)$ slots Alice is going to transmit, they can reliably exchange $O(\sqrt{n \log{T(n)}})$ bits on an AWGN channel covertly from Willie.

A critical aspect of understanding communication systems under practical constraints is their analysis and testing in finite block length regimes. While asymptotic behaviour gives us fundamental limits and important insights, it is critical to understand such systems with finite, realistic block lengths, demanded by either complexity or delay constraints. The first studies were in 
\cite{Delay_Intolerant_Covert_Communications_With_Either_Fixed_or_Random_Transmit_Power, Covert_communication_with_finite_blocklength_in_AWGN_channels, Covert_communications_with_extremely_low_power_under_finite_block_length_over_slow_fading, Delay_Constrained_Covert_Communications_With_a_Full_Duplex_Receiver, A_Finite_Block_Length_Achievability_Bound_for_Low_Probability_of_Detection_Communication, First_and_Second_Order_Asymptotics_in_Covert_Communication, Finite_Blocklength_Analysis_of_Gaussian_Random_coding_in_AWGN_Channels_under_Covert_constraints_II_Viewpoints_of_Total_Variation_Distance}. 
Needless to say, practical constrains such as limited delay and finite blocks, may limit the communication of Alice and Bob, yet, on the other hand, may limit the warden's ability to detect the communication, hence it is not a priori clear which will have a stronger effect.

Several studies considered a model in which Alice and Bob can utilize a friendly helper which
can generate Artificial Noise (AN) 
\cite{Covert_Communication_in_the_Presence_of_an_Uninformed_Jammer, Covert_Wireless_Communication_With_Artificial_Noise_Generation, Covert_Communications_with_a_Full_Duplex_Receiver_over_Wireless_Fading_Channels, Passive_Self_Interference_Suppression_for_Full_Duplex_Infrastructure_Nodes}. 
In those models, the jammer is consider either to be an ``outsider", or just Bob utilizing an additional antenna for
transmitting AN. 
Those studies have shown an improvement of the covert communication rate between Alice and Bob compared to cases without a jammer.
\subsubsection{Control}
In many contemporary applications, however, communication is used merely as means to accomplish a certain task, namely, communication is needed since the data required for the task is located on various (remote) sensors, devices or network locations. The utility in such applications depends on the task and thus, the tension when adding covertness constraints is not necessarily just between covertness and the number of bits transmitted, the task itself must be accounted for. For example, in the context of covertness when controlling a system, the tension is not just between the rate of information gathered from the system and covertness -- other aspects come into play, such as stability, delay, etc.

Focusing on control applications, it is not a-priori clear where to measure the control signal, and what are the interesting trade--offs to characterize. For example, in \cite{kostina_exact} the authors considered the minimum number of bits to stabilize a linear system, i.e., the system in \cref{eq:simple_linear_model}, where $\{ Z_{n} \}$ were assumed independent random variables, with bounded $\alpha$-th moments, and  $\{ U_{n} \}$ were the control actions, chosen by a controller, who received each time instant a single element of a finite set $\{ 1, \ldots, M \}$, as its only information about system state. The authors showed that for $|a| > 1$ (an inherently unstable system), $M = \lfloor a \rfloor + 1$ is necessary and sufficient to achieve $\beta$-moment stability, for any $\beta < \alpha$. 
Their approach was to use a normal/emergency (zoom-in/zoom-out) based controller, in order to manage the magnitude of $X_{n}$. This control signal, however, was not meant to be undetectable, and is indeed far from covert. Similar to \cite{kostina_exact},\cite{Stabilizing_a_system_with_an_unbounded_random_gain, rate-cost, Control_Over_Gaussian, Minimum_Data_Rate} 
also considered controlling a linear system, under different constraints, such as rate and energy. Yet again, covertness was not taken into account in these works.
\subsubsection{Cyber physical systems}
A closely related literature is that of security in stochastic cyber-physical systems. In this scenario, a stochastic system is being attacked, e.g., via injection of malicious data, over a communication channel. The attacker's goal is to \emph{degrade the performance} of the system while staying undetected. In particular, a relevant scenario is where a stochastic process is being controlled across a communication channel, and the control signal which is transmitted across the channel can be replaced by a malicious attacker. The controller's goal is to control the system subject to some criteria, and it may implement any arbitrary detection algorithm to detect if an attacker is present. On the other hand, the attacker's objective is to degrade the functionality of a system while staying undetected by the controller \cite{mo_secure_2009, smith_decoupled_2011, venkitasubramaniam_information-theoretic_2015}.

Mo and Sinopoli \cite{mo_false_nodate} analyzed the effects of false data injection attacks on control systems. The main components in \cite{mo_false_nodate} are a Kalman filter, in charge of estimating the system's state, and a failure detector, implemented by a quadratic form, in charge of estimating residues across the measurements in order to detect sensor failures. An attacker, who is able to inject data to a subset of the measurements, wishes to increase the state estimation error, while keeping the failure detector unalarmed. Hence, the performance of the legitimate controller is measured by the \emph{estimation error}, while the performance of the attacker is measured by the value of a quadratic form.
The authors provide a necessary and sufficient condition under which the attacker can launch a successful attack.

Recently, Bai \emph{et al.} \cite{bai_security_2015, bai_data-injection_2017} considered a related problem of stealth control in stochastic systems. Therein, a legitimate user applies a Kalman filter on the measurements of a controlled stochastic process, again, in order to estimate the system's state. An attacker can replace the control signal in the system in order to \emph{increase the estimation error of the Kalman filter}. The main contributions in \cite{bai_security_2015, bai_data-injection_2017} give converse and achievability results on the \emph{excess estimation error} of the Kalman filter as a function of the detection probability. Thus, the definition of stealthiness is different from \cite{mo_false_nodate}, yet similar to the one in this work (and in most covert communication works). However, herein, we focus on covertly \emph{controlling} a system, in the sense of changing its parameters or behavior, rather than keeping some estimation error small.

More extensions for the notion of stealthiness are discussed in \cite{zhang_stealthy_2016, kung_performance_2017}. \cite{guo_optimal_2017} studied the degradation of
remote state estimation for the case of an attacker that compromises the \emph{system measurements} based on a linear strategy. 
In \cite{weerakkody_information_2016}, the authors used a Kullback-Liebler divergence, similarly to the covert communication literature, as a measure of information flow to quantify the effect of attacks on the system output. \cite{chen_optimal_2018} characterized optimal attack strategies with respect to a linear quadratic cost that combines attackers control and undetectability goals.

Recent studies which considered the problem of controlling a stochastic cyber-physical systems that is potentially under attack, used different measures of stealthiness. For instance, the authors in \cite{fang_stealthy_2020} considered an LTI system in which the attack detector performs a hypothesis test on the innovation of the Kalman filter in order to detect malicious tampering with the actuator signals. They extended the stealthiness measure to $(\epsilon, \delta)$-stealthiness, in order to capture and quantify the degree of stealth by limiting the maximum achievable exponents of both false alarm probability and detection probability. On the other hand, the authors in \cite{an_data-driven_2018} investigated, from the attacker’s prospective, how he should design a sensor–actuator coordinated attack policy to degrade an LTI system performance while staying undetected, and without exact knowledge of the system matrices. Here the authors introduced the notion of $\alpha$-probability $\mathcal{L}_2$ stealthiness. They measure the attack’s impacts while ensuring the stealth capabilities using an $\mathcal{L}_2$-gain.

Other works considered generalized problems, such  \cite{mikhaylenko_stealthy_2021} which introduced an attack strategy called a local covert attack, which does not require access to all control input signals and sensor output signals.
The authors showed that the local covert attack can be made completely stealthy to local or global observer 
by applying the decoupling technique, i.e., manipulating all control input signals but only a part of sensor outputs.
\cite{barboni_detection_2020} on the other hand, focused on large-scale systems subject to bounded process and measurement disturbances, where a single subsystem is under a covert attack. Each subsystem can detect the presence of covert attacks in neighboring subsystems in a distributed manner. The detection strategy is based on the design of two model-based observers for each subsystem using only local information. The literature also include works on covert data leakage in controlled and cyber physical systems \cite{8353842}, or, on the other hand, works centered on covertly inserting commands to a legitimate system, yet when the target is \emph{activating a malware} installed therein, rather than changing its behaviour without being noticed \cite{8409461}. 

To conclude, while the covertness/stealthiness criteria in the studies above sometimes share similarities with the ones in this work, there are critical differences which are important to stress out. First, in this work, the objective of the legitimate controller is different: we focus on stability and system parameters, rather than estimation error. Then, we consider both converse results for \emph{any controller}, as well as direct results for specific controllers. More importantly, we identify trade--offs between the amount of information the controller uses, the covertness and the stability of the controlled system.
%
\section{Problem Formulation}
\label{sec:problem_formulation}
In the context of \cref{fig:Covert_Control} and \cref{eq:simple_linear_model}, 
\revised{this work focuses on the case where Willie is observing the system’s output.
Alice’s will is to control her system, while staying undetected by Willie.}
Alice's system is modeled as an AR(1) system, where $X_{0}$ is the initial state, $a$ is the system's \emph{gain}, $\{ Z_{n} \}$ are identically distributed, independent random variables, and $U_{n}$ is the control action at time $n$.
Choosing this kind of model stems from the fact that an AR(1) process is well studied, and can constitutes a simple model, on the one hand, yet can capture the complexity of the problem on the other. It can model systems with memory, having both random input, and a built-in system's gain. 

Here, Willie tries to detect any controlling action by Alice, observing her system's output, through some channel (clean or noisy).
Throughout this work, we assume that Willie knows the system's model, the value of $a$ and the statistical properties of $\{Z_{n}\}$, and \revised{the structure of} 
$\{U_{n}\}$. 
Thus, we can summarize Willie's hypothesis test as follows \begin{equation}\label{eq:Willies_hypothesis_test}
\begin{split}
    \mathcal{H}_{0}:& \ X_{n+1} = aX_{n} + Z_{n}
    \\
    \mathcal{H}_{1}:& \ X_{n+1} = aX_{n} + Z_{n} - U_{n}.
\end{split}
\end{equation}

The control signal, which we denote by $U_{n}$, can function in several manners. First, Alice can use it to stabilize the system, with respect to some stabilization criteria. Second, Alice can use it to alter the system's parameters or to intervene with the system's operation process.
For simplicity, we focus on controllers of the form $U_{n} = f(X_{n-1})$.

Any controller of the form $U_{n} = f(X_{n-1})$ can use different amounts of information from $X_{n-1}$ in order to control the system. 
This amount is determined by the function $f(\cdot)$. 
Explicitly, $f(\cdot)$ can use all the information in $X_{n-1}$, i.e., an infinite number of bits in the representation of $X_{n-1}$, or use some quantized value of $X_{n-1}$. In this work, besides the general result, we also consider two different specific controllers, which give a glimpse on two extreme cases. First, the case in which a minimal amount of information is taken from $X_{n-1}$, i.e., a single bit per sample. 
Second, the case in which maximal information is taken from $X_{n-1}$, i.e., an infinite number of bits. 
Those two controllers are given in 
\cref{definition:One_bit_controller} and 
\cref{definition:Threshold_controller} below, respectively.

Throughout this work, we consider two cases of system's noise, one with bounded support and the other with unbounded support.
For an AR(1) system with a bounded support noise, i.e., $\{ Z_{n }\}$ for which there is a $B > 0$, such that for any $n$, $|Z_{n}| \leq B$, we define the following controller.
\begin{definition}(\cite[equation (8)]{kostina_exact})\label{definition:One_bit_controller}
Let $U_{n}^{one}$ be the following control signal 
\begin{align}\label{eq:One_bit_controller}
U_{n}^{one} = \frac{a}{2} C_{n-1} \operatorname*{sgn}{(X_{n-1})},
\end{align}
where $a \in \mathbb{R}$, $a \neq 2$, $C_n$ is given by the following recursive formula, $C_{n} = (a/2) C_{n-1} + B$, with $C_{1} \geq \frac{B}{1- a/2}$, and $B$ is the noise bound.
We refer to $U_{n}^{one}$ as the \emph{one-bit controller}.
\end{definition}
\revised{Note that this controller keeps $X_{n}$ bounded, but only for limited support noise \cite{kostina_exact}.
It is worth mentioning that the authors in \cite{kostina_exact} use the one-bit controller in a different manner in order to deal with unbounded noise, i.e., the zoom-in zoom-out approach: using \cref{definition:One_bit_controller} in cycles in which the state process is bounded (zoom-in), but if the state process grows larger than the bound, the controller is in it's emergency mode (zoom-out) trying to catch up with the system's state to bound it again. 
Herein, we consider only the simple case in \cref{definition:One_bit_controller}. This controller is sufficient to stabilize an unstable AR(1) system with bounded noise.}
\begin{definition}\label{definition:Threshold_controller}
Let $U_{n}^{th}$ be the following control signal
\begin{align}\label{eq:Threshold_controller}
U_{n}^{th} = 
\left\lbrace
\begin{matrix}
a X_{n-1}, & |X_{n-1}| \geq D \\
0, & |X_{n-1}| < D 
\end{matrix}
\right. ,
\end{align}

where $D > 0$ is a threshold value and $a$ is the system's gain. We refer to $U_{n}^{th}$ as the \emph{threshold controller}.
\end{definition}

This intuitive controller acts as a reset to the system, that is, when the system state crosses some level, which we denote as $D$, the system resets to a zero state. 
One can generalize the controller given in \cref{definition:Threshold_controller}, to reset a system to other states, deterministic or stochastic. 
For example, resetting a system to its stationary state (if possible), as we will see next.

In order to measure the covertness of Alice's control policy, and the detectability at Willie's side, we define a covertness criterion and a detection criterion. 
Specifically, we introduce the notion of $\varepsilon$-covertness and $1-\delta$-detection.
Denote by $\alpha$ and $\beta$, the false alarm and miss detection probabilities for the hypothesis testing problem given in \cref{eq:Willies_hypothesis_test}, respectively.
\begin{definition}\label{definition:Epsilon_covertness}
We say that $\varepsilon$-covertness is achieved by Alice, if for some $\varepsilon > 0$ we have, $\alpha + \beta \geq 1 - \varepsilon$.
\end{definition}

\cref{definition:Epsilon_covertness} is a well-known criteria in covert communication, first used in  \cite{bash_limits} to establish the fundamental limit
of covert communication, and next used ubiquitously, e.g., 
\cite{Reliable_Deniable,Not_Know_Noise,Not_Know_When,dvorkind_maximizing,dvorkind_rate}. 

\begin{definition}\label{definition:1-delta_detection}
We say that $1-\delta$-detection is achieved by Willie, if for some $\delta > 0$ we have, $\alpha + \beta \leq \delta$.
\end{definition}



\subsection{Notations}
\label{subsection:Notations}
Throughout this work, we denote matrices and random vectors by capital and bold letters, such as $\textbf{A}$. Subscript are added if the dimensions are not clear from the context.
$\textbf{I}$ and $\textbf{0}$ denote the identity matrix and the all zero matrix, respectively. 
Deterministic vectors will be denoted by lowercase and bold letters, such as $\textbf{x}$.
To clarify the dimensions of the vector in question, we use a superscript, e.g., $\textbf{X}^{n}, \textbf{y}^{n}$. All the vectors are column vectors. 
We use $p$ to denote a probability mass function (PMF) of a discrete random variable or vector, and $f$ to denote a probability density function (PDF) of a continuous random variable or vector.
We denote the probability measure by $\prob{\cdot}$, the expectation operator with respect to a distribution $P$ as $\expc{P}{}{\cdot}$, similarly for the variance operator $\var{P}{}{\cdot}$. 
If the distribution is clear, we omit the subscript. 
$\log{(\cdot)}$, $\delta(\cdot)$, $u(\cdot)$ and $\indicator{\cdot}$ represent the natural logarithm, delta function, step function and the indicator function, respectively. 

\subsection{Preliminaries}
\label{subsection:Preliminary}
We introduce some relevant preliminary results and additional definitions.

We call an AR(1) process with $\{ Z_{n} \}$ being a white Gaussian noise (WGN), i.e., $\{ Z_{n} \} \sim \mathcal{N}(0, \sigma_{Z}^{2})$, a \emph{Gaussian AR(1) process}.
\begin{lemma}\label{lemma:PDF_of_n_samples_AR_1_process}
Let $\textbf{X}^{(n)} \triangleq 
\begin{pmatrix}
X_{1} & X_{2} & \cdots & X_{n}
\end{pmatrix}^{T}$ 
be an $n$-tuple of a Gaussian AR(1) process, initialized with $X_{0} \sim \mathcal{N}(0, \sigma_{0}^{2})$, independent of $\{ Z_{n} \}$ and $|a| \neq 1$. Then, the PDF of 
$\textbf{X}^{(n)}$ is given by
\begin{align}\label{eq:PDF_of_n_samples_AR_1_process}
f_{\textbf{X}^{(n)}}(\textbf{x})
=
\frac{1}{\left( 2 \pi \right)^{\frac{n}{2}} \sqrt{|\bs{\Sigma}|}} e^{-\frac{1}{2} \textbf{x}^{T} \bs{\Sigma}^{-1} \textbf{x}}, 
\end{align}
where 
$\bs{\Sigma} = \sigma_{Z}^{2} \textbf{A} \textbf{A}^{T} + \sigma_{0}^{2} \tilde{\textbf{a}} \tilde{\textbf{a}}^{T}$ is a square full-rank matrix, $[\textbf{A}]_{i,j} = a^{i-j} u(i-j)$, $[\tilde{\textbf{a}}]_{i} = a^{i}$ and $u(\cdot)$ is the discrete step function. 
I.e., $\textbf{X}^{(n)} \sim \mathcal{N}(\textbf{0}, \bs{\Sigma})$. In addition, $[\bs{\Sigma}]_{i,j} = \frac{\sigma_{Z}^{2}}{1 - a^{2}} \left( a^{|i-j|} - a^{i+j} \right) + \sigma_{0}^{2} a^{i+j}, \quad 1 \leq i,j \leq n$. 
\end{lemma}


\begin{remarkl}\label{remarkl:Stable_AR(1)_Covariance_in_SS}
Initializing the process with $\sigma_{0}^{2} = \frac{\sigma_{Z}^{2}}{1 - a^{2}}$, where $|a| < 1$, results in 
$
[\bs{\Sigma}]_{ij}
=
\frac{\sigma_{Z}^{2}}{1 - a^{2}} a^{|i-j|}, \quad 1 \leq i,j \leq n$, 
which yields a wide-sense stationary process and a stable system. \end{remarkl}
\begin{definition}(\cite[equation (2.27)]{THOMAS})
For two distributions defined over the same support, $P$ and $Q$, the Kullback-Leibler Divergence is defined by
\begin{align}\label{eq:Kullback-Leibler_divergence}
\begin{split}
\dive{KL}{P}{Q}
\triangleq
\expc{P}{}{\log{\left( \frac{P(x)}{Q(x)} \right)}}.
\end{split}
\end{align}
\end{definition}
\begin{definition}(\cite[equation (11.131)]{THOMAS}) 
For two distributions defined over the same support, $P$ and $Q$,
the total variation is defined by
\begin{align}\label{eq:Total_variation}
\begin{split}
\mathcal{V}_{T}(P,Q)
\triangleq
\frac{1}{2} \expc{Q}{}{\left\vert \frac{P(x)}{Q(x)} - 1 \right\vert}.
\end{split}
\end{align}
\end{definition}

\begin{lemma}(\cite[equation (11.138)]{THOMAS})\label{lemma:Connection_between_divergence_to_total_variation} For any two distributions defined over the same support, $P$ and $Q$, we have 
\begin{align}\label{eq:Connection_between_divergence_to_total_variation}
\begin{split}
\mathcal{V}_{T}(P,Q)
\leq
\sqrt{\frac{1}{2} \dive{KL}{P}{Q}}.
\end{split}
\end{align}
\end{lemma}
\begin{lemma}(\cite[Theorem 13.1.1]{Lehmann2005})\label{lemma:Connection_between_error_probabilities_to_total_variation}
 For the optimal test, the sum of the error probabilities, i.e., $\alpha + \beta$, is given by
\begin{align}\label{eq:Connection_between_error_probabilities_to_total_variation}
\begin{split}
\alpha + \beta
=
1 - \mathcal{V}_{T}(P,Q).
\end{split}
\end{align}
\end{lemma}
\begin{lemma}\label{lemma:Divergence_of_gaussian_vector}
Assume two multivariate normal distributions, $\mathcal{N}_{0}$ and $\mathcal{N}_{1}$, with means $\bs{\mu} _{0}$ and $\bs{\mu} _{1}$, of the same dimension $n$, and with (non-singular) covariance matrices, $\bs{\Sigma}_{0}$ and $\bs{\Sigma}_{1}$, respectively. 
Then, the Kullback–Leibler divergence between the distributions in nats is 
\begin{multline}\label{eq:Divergence_of_gaussian_vector}
D_{KL} \left( 
\mathcal{N}_{0} || \mathcal{N}_{1}
\right) = \frac{1}{2} \Bigg(
\tr{\bs{\Sigma}_{1}^{-1} \bs{\Sigma}_{0}}
+ (\bs{\mu}_{1} - \bs{\mu}_{0})^{T} \bs{\Sigma}_{1}^{-1} 
\\
(\bs{\mu}_{1} - \bs{\mu}_{0}) - n 
+ \log{\frac{|\bs{\Sigma}_{1}|}{|\bs{\Sigma}_{0}|}} \Bigg). 
\end{multline}
\end{lemma}
\revised{The proofs for \cref{lemma:PDF_of_n_samples_AR_1_process,lemma:Divergence_of_gaussian_vector} are technical, hence are given in 
\cref{appendix:1} and \cref{appendix:2}, respectively.
}.

\section{Main results}
\label{sec:main_results}
In this section, the main results are presented. First, a basic general result (\cref{theorem:An_inherently_unstable_system_cant_be_stabilized_covertly}). 
Then a negative result under a minimal–information controller (\cref{theorem:An_inherently_stable_system_cant_be_stabilized_covertly_one_bit}) and three results under the maximal-information controller  (\cref{theorem:An_inherently_stable_system_can_be_controlled_covertly,theorem:Achievable_gain_using_threshold_controller,theorem:Willie_observes_the_systems_output_for_n_samples_threshold_controller}). 
The proofs of some technical claims are given in the appendix.

We start with a general negative result. \Cref{theorem:An_inherently_unstable_system_cant_be_stabilized_covertly} states that an inherently unstable linear system, i.e., an AR(1) with $|a| > 1$, cannot be covertly stabilized. 
The stabilization criteria is absolute $\gamma$-moment stability, i.e., $\expc{}{}{|X_{n}|^{\gamma}} \leq c < \infty$. That is, stabilizing an unstable system in this context is achieving a finite absolute $\gamma$-moment for the system's state. 
\begin{theorem}\label{theorem:An_inherently_unstable_system_cant_be_stabilized_covertly}
Consider the linear stochastic system in \cref{eq:simple_linear_model}, with $|a| > 1$, an i.i.d.\ $Z_{n}$ and a control signal $U_{n}$ which keeps the system $\gamma$-moment stable. 
If Willie knows a uniform bound on the system's $\gamma$-moment, i.e., $\expc{}{}{|X_{n}|^{\gamma}} \leq c < \infty$ and observes the system's output through a clean channel, he can achieve $1-\delta$-detection for any $\delta > 0$. For a Gaussian AR(1), if Willie observes the system's output at a time $n_{0}$, where 
\[
n_{0}
\geq
\frac{\log{\frac{M\sqrt{a^{2} - 1}}{\sigma_{Z} Q^{-1}\left( \frac{1 - \frac{\delta}{2}}{2} \right)}}}
{\log{|a|}}
\]
and $M \geq \sqrt[\gamma]{\frac{2c}{\delta}}$, he can achieve $1-\delta$-detection.
\end{theorem}


\begin{proof}
An AR(1) process with $|a| > 1$ is not stable, in the sense that $\expc{}{}{X_{n}^{2}} \xrightarrow[n \to \infty]{} \infty$ \cite{Time_Seires_Analysis}. 
Thus, Willie can observe $X_{n_{0}}$ and compare $|X_{n_{0}}|$ with some large constant $M$, which depends on $c$. Based on his observation, he decides between the following two hypotheses in \cref{eq:Willies_hypothesis_test}, 
where $U_{n}$ is such that $\expc{}{}{|X_{n_{0}}|^{\gamma}} \leq c$. 
We bound the miss detection probability as follows, 
\begin{align*}
\beta 
=&
\prob{|X_{n_{0}}| > M | \mathcal{H}_{1}}
\\ \stackrel{(a)}{\leq}&
\frac{\expc{}{}{|X_{n_{0}}|^{\gamma} | \mathcal{H}_{1}}}{M^{\gamma}}
\\ \stackrel{(b)}{\leq}&
\frac{c}{M^{\gamma}}
\\ \stackrel{(c)}{\leq}&
\frac{\delta}{2},
\end{align*}

where (a) is due to Markov's inequality, (b) since $U_{n}$ stabilizes the system in the sense that $\expc{}{}{|X_{n_{0}}|^{\gamma} | \mathcal{H}_{1}} \leq c < \infty$. 
(c) is when Willie sets $M \geq \sqrt[\gamma]{\frac{2c}{\delta}}$ to get $\beta \leq \frac{\delta}{2}$. 

On the other hand, 
we evaluate the false alarm probability as follows, 
\begin{align*}
\alpha
=&
\prob{|X_{n_{0}}| \leq M | \mathcal{H}_{0}}
\\=&
\prob{\left\vert \sum_{k=1}^{n_{0}}{a^{n_{0}-k} Z_{k}} \right\vert < M}
\\\stackrel{}{=}&
\prob{- \frac{M}{|a|^{n_{0}}} < \sum_{k=1}^{n_{0}}{a^{-k} Z_{k}} < \frac{M}{|a|^{n_{0}}}}.
\end{align*}

For $|a| > 1$, the sum $\tilde{Z} = \sum_{k=1}^{n_{0}}{a^{-k} Z_{k}}$ has a variance of $\var{}{}{\tilde{Z}} = \sigma_{Z}^{2} \frac{1 - a^{-2n_{0}}}{a^{2} - 1}$, which is non-zero for any $n_{0}$. 
On the other hand, $\frac{M}{|a|^{n_{0}}} \xrightarrow[n_{0} \to \infty]{} 0$, hence $\alpha \xrightarrow[n_{0} \to \infty]{} 0$. 

In the special case in which $Z_{k} \sim \mathcal{N}(0,\sigma_{Z}^{2})$, we have 
\begin{align*}
\alpha
=&
\prob{- \frac{M}{|a|^{n_{0}}} < \sum_{k=1}^{n_{0}}{a^{-k} Z_{k}} < \frac{M}{|a|^{n_{0}}}}
\\=&
1 - 2Q\left( \frac{M\sqrt{a^{2} - 1}}{\sigma_{Z} \sqrt{a^{2n} - 1}} \right)
\\ \stackrel{(a)}{\leq}&
\frac{\delta}{2}
\end{align*}

where (a) is by applying the detection constraint. Rearranging terms yields the requirement.
\end{proof}


\subsection{Negative result under a minimal-information controller}\label{subsetion:Converse_results_under_a_minimal-information_controller}

\revised{\Cref{theorem:An_inherently_stable_system_cant_be_stabilized_covertly_one_bit} introduces the case in which Willie observes Alice's system's output via a clean channel, while Alice uses the one-bit controller. 
The one-bit controller is known to stabilize an unstable system with bounded noise \cite{kostina_exact}, however, \cref{theorem:An_inherently_unstable_system_cant_be_stabilized_covertly} 
asserts that covert control cannot be achieved.  
Albeit that, one can consider using the one-bit controller in an inherently stable system, i.e., $|a| < 1$, in order to restrain the system's state within a strict bound, for example.
Still, \cref{theorem:An_inherently_stable_system_cant_be_stabilized_covertly_one_bit} 
asserts that Willie can achieve $1-\delta$-detection, for any $\delta > 0$, as the number of observations increases.}
\begin{theorem}
\label{theorem:An_inherently_stable_system_cant_be_stabilized_covertly_one_bit}
Consider an AR(1) system with $|a| < 1$, and $U_{n}$ as the one-bit controller. 
If Willie observes the system's output through a clean channel for at least
\small
\[
K_{0}(\delta)
=
\frac{1}{E_{U}^{2}} \left( \sqrt{\frac{m_{Z}(4) - \sigma_{Z}^{4} + 4E_{U} \sigma_{Z}^{2}}{\delta/2}}
+
\sqrt{\frac{m_{Z}(4) - \sigma_{Z}^{4}}{\delta/2}}
\right)^{2}
\]
\normalsize

time samples, where $m_{Z}(4) \triangleq \expc{}{}{Z^{4}}$, $\sigma_{Z}^{2} \triangleq \var{}{}{Z} = \expc{}{}{Z^{2}}$,
$E_{U} \triangleq \left( \frac{aB}{2-a} \right)^{2}$ and $B$ is some deterministic number for which $|Z| \leq B$, 
for any $\delta > 0$, Willie achieves $1-\delta$-detection.
\end{theorem}


\revised{Note, however, that this result is under a minimal information controller, limiting Alice’s ability to gather enough information before each control decision. The result in the next subsection will show that with more information, Alice can do better and stay undetected.}
In order to prove \cref{theorem:An_inherently_stable_system_cant_be_stabilized_covertly_one_bit}, we first bound the energy of the one-bit controller. \cref{claim:One_bit_controller_energy} give an upper and lower bounds for the energy of the control signal shown in \cref{eq:One_bit_controller}.
\begin{claim}\label{claim:One_bit_controller_energy}
The energy of the controller in  \cref{eq:One_bit_controller} satisfies
\begin{align}\label{eq:One_bit_controller_energy}
\begin{split}
\left( \frac{a B}{2 - a} \right)^{2} \leq E_{U} \leq \left( \frac{a}{2} C_{1} \right)^{2},
\end{split}
\end{align}
 
where 
$C_{n}$ is a deterministic, monotonically decreasing series, converging to $\frac{B}{1 - a/2}$. Moreover, 
$C_{1}$ can be arbitrary number which sustains $C_{1} \geq \frac{B}{1- a/2}$, and $C_{n} = (a/2) C_{n-1} + B$, where $B$ is the noise bound, and $a \in \mathbb{R},  a \neq 2$ (see \cref{definition:One_bit_controller}).
\end{claim}

\begin{remarkc}\label{remarkc:One_bit_controller_energy_steady_state}
Since $C_{1}$ can be arbitrary number which sustains $C_{1} \geq \frac{B}{1- a/2}$, 
for simplicity, we set $C_{1} = \frac{B}{1 - a/2}$ unless otherwise stated, thus,  
\begin{align}\label{eq:One_bit_controller_energy_steady_state}
\begin{split}
E_{U} = \left( \frac{a B}{2 - a} \right)^{2},
\end{split}
\end{align}

which yields a constant energy with respect to time.
Furthermore, 
the above also holds as $N \to \infty$ regardless of the choice of $C_{1}$ (however, $C_{1}$ has to be greater or equal to $\frac{B}{1 - a/2}$). 
Hence, the average energy of \cref{eq:One_bit_controller} in steady state, is given by \cref{eq:One_bit_controller_energy_steady_state}. 
\end{remarkc}

The proofs are given in the \cref{appendix:3}.

\begin{proof}
Willie is observing to Alice's system's output through a clean channel, when $|a| < 1$ and in steady state. Therefore, Willie's observation at time $n$, is an AR(1) signal controlled or not.
We prove the impossibility of covert control in this case, using the following detection method: 
Willie observes the system's output through a clean channel for $K+1$ samples, in any time sample $n$ Willie evaluates $Y_{n} \triangleq X_{n} - aX_{n-1}$, and then calculates the average energy of $Y_{n}$, i.e., 
$E_{W} = \frac{1}{K}\sum_{n}^{}{Y_{n}^{2}}$. 
Willie compares $E_{W}$ to some expected energy level, in order to decide if the system is being controlled or not.

Willie preforms hypotheses testing approach to decide if Alice is controlling the system or not, he uses the following hypotheses,
\begin{align*}
\begin{matrix}
\mathcal{H}_{0}: &
Y_{n} = Z_{n}, & 
n = k, \ldots, K + k - 1,
\\
\mathcal{H}_{1}: &
Y_{n} = Z_{n} - U_{n}, &
n = k, \ldots, K + k - 1.
\end{matrix}
\end{align*}

Under the null hypothesis, Willie observes an i.i.d.\ process, thus, the mean and the variance of $E_{W}$ under the assumption that $\mathcal{H}_{0}$ is true are,
\begin{align}\label{eq:One_bit:Output:Mean_H_0}
\begin{split}
\mathbb{E} \left[ E_{W}| \mathcal{H}_{0} \right]
=&
\mathbb{E} \left[ \frac{1}{K}\sum_{n}^{}{ Y_{n}^{2}} | \mathcal{H}_{0} \right]
\\=&
\frac{1}{K}\sum_{n}^{}{ \mathbb{E} \left[ Z_{n}^{2}\right]} 
\\=&
\sigma_{Z}^{2},
\end{split}
\end{align}
\begin{align}\label{eq:One_bit:Output:Variance_H_0}
\begin{split}
\var{}{}{ E_{W}| \mathcal{H}_{0} }
=&
\var{}{}{ \frac{1}{K}\sum_{n}^{}{ Y_{n}^{2}}| \mathcal{H}_{0}}
\\=&
\frac{1}{K^{2}} \sum_{n}^{}{ \var{}{}{ Z_{n}^{2}}}
\\=&
\frac{1}{K^{2}} \sum_{n}^{}{ (m_{Z}(4) - \sigma_{Z}^{4})}
\\=&
\frac{m_{Z}(4) - \sigma_{Z}^{4}}{K},
\end{split}
\end{align}

where $m_{Z}(4)$ is the fourth moment of $Z$. 
In a similar fashion, under the alternative hypothesis, Willie observes an i.i.d.\ process with the control signal, which are both independent at the same time samples, thus, the mean and the variance of $E_{W}$ under the assumption that $\mathcal{H}_{1}$ is true are,
\begin{align}\label{eq:One_bit:Output:Mean_H_1}
\begin{split}
\mathbb{E} \left[ E_{W}| \mathcal{H}_{1} \right]
=&
\mathbb{E} \left[ \frac{1}{K}\sum_{n}^{}{ Y_{n}^{2}} | \mathcal{H}_{1} \right]
\\=&
\frac{1}{K}\sum_{n}^{}{ \mathbb{E} \left[ (Z_{n} - U_{n})^{2}\right]} 
\\=&
\frac{1}{K}\sum_{n}^{}{ (\expc{}{}{Z_{n}^{2}} + \expc{}{}{U_{n}^{2}})} 
\\=&
\sigma_{Z}^{2} + E_{U},
\end{split}
\end{align}

where $Z_{n} \indep U_{n}$ and $U_{n}^{2} = E_{U}$. 
\begin{align}\label{eq:One_bit:Output:Second_moment_H_1}
\begin{split}
\expc{}{}{ E_{W}^{2}| \mathcal{H}_{1} }
=&
\frac{1}{K^{2}} 
\sum_{n,m}^{}{ \expc{}{}{(Z_{n} - U_{n})^{2} (Z_{m} - U_{m})^{2}}}
\\\stackrel{(a)}{=}&
\frac{1}{K^{2}} \left( 
\sigma_{Z}^{4} K (K-1) + m_{Z}(4) K \right.
\\&+ \left.
2 E_{U} \sigma_{Z}^{2} K^{2} 
+
4 E_{U} \sigma_{Z}^{2} K + E_{U}^{2} K^{2}
\right)
\\=&
\left( \sigma_{Z}^{2} + E_{U} \right)^{2} 
\\&+ 
\frac{m_{Z}(4) - \sigma_{Z}^{4} + 4E_{U} \sigma_{Z}^{2}}{K},
\end{split}
\end{align}

where (a) is due to, 
\begin{align*}
\small
\mathbb{E} &\left[ (Z_{n} - U_{n})^{2} (Z_{m} - U_{m})^{2} \right]
\\=&
\expc{}{}{(Z_{n}^{2} - 2 Z_{n} U_{n} + U_{n}^{2}) (Z_{m}^{2} - 2 Z_{m} U_{m} + U_{m}^{2})}
\\=&
\mathbb{E}\left[ Z_{n}^{2} Z_{m}^{2} - 2 Z_{n}^{2} Z_{m} U_{m} + Z_{n}^{2} U_{m}^{2}
\right.
\\&
-2 Z_{m}^{2} Z_{n} U_{n} + 4 Z_{n} U_{n} Z_{m} U_{m} - 2 Z_{n} U_{n} U_{m}^{2}
\\&
\left.
+ Z_{m}^{2} U_{n}^{2} - 2 Z_{m} U_{m} U_{n}^{2} + U_{n}^{2} U_{m}^{2}
\right]
\\=&
m_{Z}(4) \delta(n-m) + \sigma_{Z}^{4} (1 - \delta(n-m)) + E_{U} \sigma_{Z}^{2}
\\&
+ 4 E_{U} \sigma_{Z}^{2} \delta(n-m) 
\\&+
4 \expc{}{}{Z_{n} U_{n} Z_{m} U_{m}} (1 - \delta(n-m))
\\&+
E_{U} \sigma_{Z}^{2} + E_{U}^{2}
\\\stackrel{(b)}{=}&
m_{Z}(4) \delta(n-m) + \sigma_{Z}^{4} (1 - \delta(n-m)) + 2 E_{U} \sigma_{Z}^{2} 
\\& 
+ 4 E_{U} \sigma_{Z}^{2} \delta(n-m) + E_{U}^{2},
\normalsize
\end{align*}
and (b) above is since
\begin{align*}
\small
\mathbb{E} & \left[ Z_{n} U_{n} Z_{m} U_{m}  \right] (1 - \delta(n-m))
\\=&
1_{m < n} \expc{}{}{Z_{n}} \expc{}{}{U_{n} Z_{m} U_{m}}
+
1_{m > n} \expc{}{}{Z_{m}} \expc{}{}{Z_{n} U_{n} U_{m}}
\\=&
0.
\normalsize
\end{align*}
Summing the expression above yields
\begin{align*}
&\sum_{n,m}^{}{ \expc{}{}{(Z_{n} - U_{n})^{2} (Z_{m} - U_{m})^{2}}} = 
\\=&
m_{Z}(4) K 
+
\sigma_{Z}^{4} K (K - 1) 
+
2 E_{U} \sigma_{Z}^{2} K^{2} 
+
4 E_{U} \sigma_{Z}^{2} K 
\\&+
E_{U}^{2} K^{2}.
\end{align*}
Using \cref{eq:One_bit:Output:Mean_H_1,eq:One_bit:Output:Second_moment_H_1}, 
\begin{align}\label{eq:One_bit:Output:Variance_H_1}
\begin{split}
\var{}{}{ E_{W}| \mathcal{H}_{1} }
=&
\expc{}{}{ E_{W}^{2}| \mathcal{H}_{1} } - \expc{}{2}{ E_{W}| \mathcal{H}_{1} }
\\\stackrel{}{=}&
\frac{m_{Z}(4) - \sigma_{Z}^{4} + 4E_{U} \sigma_{Z}^{2}}{K}.
\end{split}
\end{align}
If $\mathcal{H}_{0}$ is true, then $E_{W}$ should be close to $\mathbb{E}\left[ E_{W}| \mathcal{H}_{0} \right]$. Willie picks a threshold which we denote as $t$, and compares the value of $E_{W}$ to $\sigma_{Z}^{2} + t$. Willie accepts $\mathcal{H}_{0}$ if  $E_{W} < \sigma_{Z}^{2} + t$ and rejects it otherwise. 
We bound the false alarm probability using 
\cref{eq:One_bit:Output:Variance_H_0,eq:One_bit:Output:Mean_H_0} and with Chebyshev’s inequality,
\begin{align*}
\begin{split}
\alpha
=&
\mathbb{P}\{ E_{W} \geq \sigma_{Z}^{2} + t| \mathcal{H}_{0} \}
\\\leq&
\mathbb{P}\{ |E_{W} - \sigma_{Z}^{2}| \geq t| \mathcal{H}_{0} \}
\\\leq&
\frac{m_{Z}(4) - \sigma_{Z}^{4}}{K t^{2}}.
\end{split}
\end{align*}

Thus, to obtain $\alpha \leq \frac{\delta}{2}$, Willie sets $t = \sqrt{\frac{m_{Z}(4) - \sigma_{Z}^{4}}{K \delta / 2}}$.
The probability of a miss detection, $\beta$, is the probability that $E_{W} < \sigma_{Z}^{2} + t$ when $\mathcal{H}_{1}$ is true. We bound $\beta$ using \cref{eq:One_bit:Output:Variance_H_1,eq:One_bit:Output:Mean_H_1} and with Chebyshev’s inequality,
\begin{align*}
\begin{split}
\beta
=&
\mathbb{P}\{ E_{W} < \sigma_{Z}^{2} + t| \mathcal{H}_{1} \}
\\\leq&
\mathbb{P}\{ |E_{W} - (\sigma_{Z}^{2} + E_{U})| > E_{U} - t| \mathcal{H}_{1} \}
\\\leq&
\frac{m_{Z}(4) - \sigma_{Z}^{4} + 4E_{U} \sigma_{Z}^{2}}{K (E_{U} - t)^{2}}
\\=&
\frac{m_{Z}(4) - \sigma_{Z}^{4} + 4E_{U} \sigma_{Z}^{2}}{\left(\sqrt{K} E_{U} - \sqrt{\frac{m_{Z}(4) - \sigma_{Z}^{4}}{\delta / 2}} \right)^{2}}.
\end{split}
\end{align*}

Thus, to obtain $\beta \leq \frac{\delta}{2}$, Willie sets his observation window to be at least
\small
\[
K_{0}(\delta)
=
\frac{1}{E_{U}^{2}} \left( \sqrt{\frac{m_{Z}(4) - \sigma_{Z}^{4} + 4E_{U} \sigma_{Z}^{2}}{\delta/2}}
+
\sqrt{\frac{m_{Z}(4) - \sigma_{Z}^{4}}{\delta/2}}
\right)^{2}.
\]
\normalsize
By doing so, Willie can detect with arbitrarily low error probability Alice's control actions with the one-bit controller, i.e., Willie achieves $1-\delta$-detection for any $\delta > 0$.
\end{proof}

\subsection{Results under a maximal-information controller}
\label{subsetion:Results_under_a_maximal-information_controller}

\Cref{theorem:An_inherently_stable_system_can_be_controlled_covertly,theorem:Achievable_gain_using_threshold_controller} 
below state that an inherently stable linear system can be covertly controlled. 
Herein $|a| < 1$, hence the system is already stable, thus, a stabilization action is not needed. 
However, if Alice desires to alter or interfere with the system's operation, it is possible to do so while keeping Willie ignorant about these actions.

\Cref{theorem:An_inherently_stable_system_can_be_controlled_covertly} considers the case in which Alice desires to change the gain of the system, i.e., her goal is to change an AR(1) system with a gain of $a$, to an AR(1) system with a gain of $b$. 

\begin{theorem}\label{theorem:An_inherently_stable_system_can_be_controlled_covertly}
Consider the system given in \cref{eq:simple_linear_model} for $|a| < 1$, in steady state, $Z_{n} \sim \mathcal{N}(0, \sigma_{Z}^{2})$ i.i.d.  and $U_{n} = (a-b) X_{n-1}$, where $0 < |a| < |b| < 1$ and $\sgn{a} = \sgn{b}$. 
If Willie observes the system's output through a clean channel, for a time window $n < \frac{2b}{b-a}$, knows the structure of $U_{n}$ and $b$ satisfies $|b|<\sqrt{1 - (1-a^{2}) e^{-4 \epsilon^{2}}}$, 
then for any method of detection that Willie uses, Alice achieves an $\epsilon$-covertness for any $\epsilon > 0$.
\end{theorem}

In other words, \cref{theorem:An_inherently_stable_system_can_be_controlled_covertly} states that with no information constraint on Alice's behalf, an inherently stable AR(1) system can be covertly controlled, in the sense that Alice can change the system's gain, $a$, to a different one, $b$, without being detected (to some covertness level) by Willie.
In addition, a large change in the system’s gain by Alice, while staying covert, is possible when Willie is restricted to a smaller observation window.
\cref{theorem:An_inherently_stable_system_can_be_controlled_covertly} gives us a look at the best case that Willie can have (a clean channel), and still assures that Willie can not detect Alice's control actions. 

\begin{proof}
On Willie's side, we have the following hypothesis testing problem,  
\begin{align*}
\begin{matrix}
\mathcal{H}_{0}: &
X_{n} = a X_{n-1} + Z_{n}, & 
n \geq 1,
\\
\mathcal{H}_{1}: &
X_{n} = b X_{n-1} + Z_{n}, & 
n \geq 1 .
\end{matrix}
\end{align*}

Since $Z_{n} \sim \mathcal{N}(0, \sigma_{Z}^{2})$ i.i.d., then $\textbf{X}^{(n)}| \mathcal{H}_{0} \sim \mathcal{N} \left( \textbf{0}, \bs{\Sigma_{0}} \right)$ and
$\textbf{X}^{(n)}| \mathcal{H}_{1} \sim \mathcal{N} \left( \textbf{0}, \bs{\Sigma_{1}} \right)$, where $[\bs{\Sigma_{0}}]_{i,j} = \frac{\sigma_{Z}^{2}}{1 - a^{2}} a^{|i-j|}$ and $[\bs{\Sigma_{1}}]_{i,j} = \frac{\sigma_{Z}^{2}}{1 - b^{2}} b^{|i-j|}$ for $1 \leq i,j \leq n$, respectively 
(see \cref{lemma:PDF_of_n_samples_AR_1_process} and \cref{remarkl:Stable_AR(1)_Covariance_in_SS}). Hence, 
\begin{align*}
\alpha + \beta
\stackrel{(a)}{=}&
1 - 
\mathcal{V}_{T}\left( f_{\textbf{X}| \mathcal{H}_{0}}, f_{\textbf{X}| \mathcal{H}_{1}} \right)
\\ \stackrel{(b)}{\geq}&
1 - 
\sqrt{\frac{1}{2} \dive{KL}{f_{\textbf{X}| \mathcal{H}_{0}}}{f_{\textbf{X}| \mathcal{H}_{1}}}}
\\ \stackrel{(c)}{=}&
1 - 
\sqrt{\frac{1}{4} \left( \tr{\bs{\Sigma}_{1}^{-1} \bs{\Sigma}_{0}} - n + \log{\left( \frac{\left\vert  \bs{\Sigma}_{1} \right\vert}{\left\vert  \bs{\Sigma}_{0} \right\vert} \right)} \right)}
\\ \stackrel{(d)}{>}&
1 - \frac{1}{2} 
\sqrt{\log{\left( \frac{\left\vert  \bs{\Sigma}_{1} \right\vert}{\left\vert  \bs{\Sigma}_{0} \right\vert} \right)}}
\\ \stackrel{(e)}{>}&
1 - \frac{1}{2} 
\sqrt{\log{\left( \frac{1 - a^{2}}{1 - b^{2}} \right)}}
\\ \stackrel{(f)}{>}&
1 - 
\varepsilon,
\end{align*}

where (a) 
is due to 
\cref{lemma:Connection_between_error_probabilities_to_total_variation}, (b) is due to 
\cref{lemma:Connection_between_divergence_to_total_variation} and (c) is by \cref{lemma:Divergence_of_gaussian_vector}.  
(d) is 
by \cref{claim:trace_B_inv_A} below.
(see \cref{appendix:4} for the proof).
\begin{claim}\label{claim:trace_B_inv_A}
Consider two $n$-tuples of Gaussian AR(1) processes in steady state. 
The first with a gain $a$ and a covariance matrix $\bs{\Sigma}_{0}$, and the second with a gain $b$ and a covariance matrix $\bs{\Sigma}_{1}$. 
For $|a|, |b| < 1$, we have
\begin{align*}
\tr{\bs{\Sigma}_{1}^{-1} \bs{\Sigma}_{0}}
=&
\frac{(n-2) b^{2} - 2(n-1) ab + n}{1 - a^{2}},
\end{align*}
\end{claim}
hence,
\begin{align*}
\tr{\bs{\Sigma}_{1}^{-1} \bs{\Sigma}_{0}} - n
=&
\frac{(b-a)^{2}}{1 - a^{2}}\left(
n  - \frac{2b}{b-a}
\right)
<
0.
\end{align*}

Finally, (e) is by the substitution of $\left\vert  \bs{\Sigma}_{0} \right\vert = \frac{\sigma_{Z}^{2n}}{1 - a^{2}}$ and $\left\vert  \bs{\Sigma}_{1} \right\vert = \frac{\sigma_{Z}^{2n}}{1 - b^{2}}$ (see \cref{appendix:4} for the proof of \cref{claim:Divergence_of_controlled_and_uncontrolled_AR_1_of_stable_system_in_steady_state}) and 
(f) is by applying the covertness criterion. 
This results in 
\[
\varepsilon
>
\frac{1}{2} \sqrt{\log{\left( \frac{1 - a^{2}}{1 - b^{2}} \right)}}.
\]
\end{proof}

Consider now another control objective, which is resetting the system's memory.  
\Cref{theorem:Achievable_gain_using_threshold_controller} introduces an achievable range of gains of an AR(1) system, for which Alice can reset the system to its stationary distribution while staying undetected by Willie. 
It is restricted, however, to the case of one control action in the system's operation time, and under steady-state conditions. 
On the other hand, Willie is observing the system's output through a clean channel, for the whole of the system's operation time, and he is unrestricted in terms of complexity or strategy used. 
\begin{theorem}\label{theorem:Achievable_gain_using_threshold_controller}
Consider a Gaussian AR(1) system with $|a| < 1$. Alice is using the threshold controller, and Willie is observing the system's output through a clean channel. 
If the system is being controlled by resetting at one time sample $1 \leq \tau < N$, to its stationary distribution, i.e., $X_{\tau + 1} = a \tilde{X} + Z_{\tau + 1}$, where $\tilde{X} \sim \mathcal{N}\left(0, \frac{\sigma_{Z}^{2}}{1 - a^{2}} \right)$ and the system's gain, $a$, satisfies $|a| 
\leq 
\sqrt{1 - e^{-4 \varepsilon^{2}}}$, then for any method of detection that Willie will use,  Alice achieves $\varepsilon$-covertness for any $\varepsilon > 0$. 
\end{theorem}

The fact that Willie observes the system's output through a clean channel, gives the bound in \cref{theorem:Achievable_gain_using_threshold_controller} a fundamental value. 
Furthermore, the bound given in \cref{theorem:Achievable_gain_using_threshold_controller} 
holds even in the extreme case, in which Willie knows the potential reset time. 

To prove \cref{theorem:Achievable_gain_using_threshold_controller}, we will use \cref{lemma:Connection_between_error_probabilities_to_total_variation,lemma:Connection_between_divergence_to_total_variation} to bound the sum of the error probabilities. Yet, to do so, we first need several supporting claims, to upper bound the relevant Kullback-Leibler divergence.
This is done in the following steps.
Consider a specific case of a Gaussian AR(1) system, with and without the threshold controller. In this case, 
given $X_{k-1} = x_{k-1}$, $X_{k} \sim \mathcal{N}(a x_{k-1}, \sigma_{Z}^{2})$, $\forall 1 \leq k \leq \tau$, where $\tau$ is the first threshold crossing time, and $X_{0}$ is distributed according to the steady-state distribution (see \cref{remarkl:Stable_AR(1)_Covariance_in_SS}). 

Denote $\textbf{X}^{(n)} \triangleq [X_{1}, X_{2}, \ldots, X_{n}]^{T}$. 
For $n \leq \tau$, i.e., without any control action, the probability density function of an AR(1) process, can be easily evaluated (see \cref{lemma:PDF_of_n_samples_AR_1_process}). 
If there is only one crossing time, which we indicate as $1 \leq \tau < n$, then, $\textbf{X}^{(1,\tau)} | {\tau} = [X_{1}, \ldots, X_{\tau}]^{T}$ and $\textbf{X}^{(\tau + 1,n)} | {\tau} = [X_{\tau + 1}, \ldots, X_{n}]^{T}$ are two independent random vectors. 
Thus, we have the following. 
\begin{fact}\label{fact:Cond_PDF_of_controlled_AR_1_process}
Consider a Gaussian AR(1) system. When using the threshold controller, the PDF of the system state vector $\tilde{\textbf{X}}^{(n)}$ conditioned on $\tau$, when there is only one crossing time, is given by
\begin{align}\label{eq:Cond_PDF_of_controlled_AR_1_process}
\begin{split}
f_{\tilde{\textbf{X}}^{(n)} |\tau}(\textbf{x} | \tau)
=
\frac{1}{\left( 2 \pi \right)^{\frac{n}{2}} \sqrt{|\tilde{\bs{\Sigma}}_{n}|}} e^{-\frac{1}{2} \textbf{x}^{T} \tilde{\bs{\Sigma}}_{n}^{-1} \textbf{x}},
\end{split}
\end{align}

with $\tilde{\bs{\Sigma}}_{n}$ given by 
\begin{align*}
\tilde{\bs{\Sigma}}_{n}
=&
\left(
\begin{matrix}
\bs{\Sigma}_{\tau} & \textbf{0} \\
\textbf{0} & \bs{\Sigma}_{n - \tau}
\end{matrix}
\right).
\end{align*}

I.e., $\tilde{\textbf{X}}^{(n)} \sim \mathcal{N}(\textbf{0}, \tilde{\bs{\Sigma}}_{n})$ and $\bs{\Sigma}_{k}$ is the covariance matrix of $k$ samples Gaussian AR(1) process.
\end{fact}

We can now turn to the main technical claim.
\begin{claim}\label{claim:Divergence_of_controlled_and_uncontrolled_AR_1}
Let $\textbf{X}^{(n)}$ be the vector of system states for an uncontrolled Gaussian AR(1). Denote by $\tilde{\textbf{X}}^{(n)}$ the vector of system states under one control action. 
Then,
\begin{align}\label{eq:Divergence_of_controlled_and_uncontrolled_AR_1}
\begin{split}
D_{KL} & \left( 
f_{\textbf{X}^{(n)}}(\textbf{x}) || f_{\tilde{\textbf{X}}^{(n)}}(\textbf{x})
\right)
\\\leq&
\expc{p_{\tau}}{}{
\dive{KL}{f_{\textbf{X}^{(n)}}(\textbf{x})} {f_{\tilde{\textbf{X}}^{(n)} | \tau}(\textbf{x} | \tau  = \tau_{1})}},
\end{split}
\end{align}

where $\bs{\Sigma}_{\tilde{\textbf{X}}^{(n)} |\tau}$ and $\bs{\Sigma}_{\textbf{X}^{(n)}}$ are the covariance matrices of $\tilde{\textbf{X}}^{(n)}  | {\tau}$ and $\textbf{X}^{(n)}$, respectively. 
\end{claim}

\begin{proof}
We have,
\begin{align*}
D_{KL} &\left( 
f_{\textbf{X}^{(n)}} || f_{\tilde{\textbf{X}}^{(n)}}
\right)
\\=&
\expc{f_{\textbf{X}^{(n)}}}{}{
\log{f_{\textbf{X}^{(n)}}} - \log{\sum_{\tau_{1} = 1}^{n}{f_{\tilde{\textbf{X}}^{(n)} | \tau}(\textbf{x} | \tau  = \tau_{1})
p_{\tau}}}}
\\\stackrel{}{=}&
-H(f_{\textbf{X}^{(n)}})
- \expc{f_{\textbf{X}^{(n)}}}{}{\log{ \expc{p_{\tau}}{}{f_{\tilde{\textbf{X}}^{(n)} | \tau}(\textbf{x} | \tau  = \tau_{1})}}}
\\\stackrel{(a)}{\leq}&
-H(f_{\textbf{X}^{(n)}})
- \expc{f_{\textbf{X}^{(n)}}}{}{\expc{p_{\tau}}{}{\log{ f_{\tilde{\textbf{X}}^{(n)} | \tau}(\textbf{x} | \tau  = \tau_{1})}}}
\\\stackrel{(b)}{=}&
-H(f_{\textbf{X}^{(n)}})
- \expc{p_{\tau}}{}{\expc{f_{\textbf{X}^{(n)}}}{}{\log{ f_{\tilde{\textbf{X}}^{(n)} | \tau}(\textbf{x} | \tau  = \tau_{1})}}}
\\\stackrel{(c)}{=}&
\expc{p_{\tau}}{}{H_{CE} \left( f_{\textbf{X}^{(n)}}(\textbf{x}), f_{\tilde{\textbf{X}}^{(n)} | \tau}(\textbf{x} | \tau  = \tau_{1}) \right)}
-H(f_{\textbf{X}^{(n)}})
\\\stackrel{(d)}{=}&
\expc{p_{\tau}}{}{
\dive{KL}{f_{\textbf{X}^{(n)}}(\textbf{x})} {f_{\tilde{\textbf{X}}^{(n)} | \tau}(\textbf{x} | \tau  = \tau_{1})}}
\end{align*}

(a) is due to Jensen's inequality. 
(b) is due to changing the order of the expectations, since $\expc{p_{\tau}}{}{\cdot}$ is a discrete and finite expectation. 
(c) is by the definition of the \emph{cross-entropy}, 
$H_{CE}(P,Q) = -\expc{P}{}{\log{Q}}$. 
(d) is due to the following relation: $\dive{KL}{P}{Q} = H_{CE}(P,Q) - H(P)$, and since $H(f_{\textbf{X}^{(n)}})$ is independent of $\tau$. 
\end{proof}

\begin{claim}\label{claim:Divergence_of_controlled_and_uncontrolled_AR_1_of_stable_system_in_steady_state}
Consider a Gaussian AR(1) system with $|a| < 1$, and the system is in steady state, then
\begin{align*}
\begin{split}
\dive{KL}{f_{\textbf{X}^{(n)}}(\textbf{x})} {f_{\tilde{\textbf{X}}^{(n)} | \tau}(\textbf{x} | \tau)}
=&
\frac{1}{2} \log{\frac{1}{1 - a^{2}}}.
\end{split}
\end{align*}
\end{claim}

\begin{corollaryc}\label{corollaryc:Diveragence_upper_bound}
For a Gaussian AR(1) system with $|a| < 1$ and in steady state, by 
\cref{claim:Divergence_of_controlled_and_uncontrolled_AR_1_of_stable_system_in_steady_state,claim:Divergence_of_controlled_and_uncontrolled_AR_1},
we get, 
\begin{align*}
\dive{KL}{f_{\textbf{X}^{(n)}}}{f_{\tilde{\textbf{X}}^{(n)}}}
\leq 
\frac{1}{2} \log
\frac{1}{1 - a^{2}}, 
\end{align*}

since 
$\dive{KL}{f_{\textbf{X}^{(n)}}(\textbf{x})} {f_{\tilde{\textbf{X}}^{(n)} | \tau}(\textbf{x} | \tau)},$ 
does not depend on $\tau$.
\end{corollaryc}

The proof of  \cref{claim:Divergence_of_controlled_and_uncontrolled_AR_1_of_stable_system_in_steady_state} is given in
\cref{appendix:5}.

We can now give the proof of  \cref{theorem:Achievable_gain_using_threshold_controller}.

\begin{proof}(\cref{theorem:Achievable_gain_using_threshold_controller})
We know that, 
\begin{align*}
\alpha + \beta
\stackrel{(a)}{=}&
1 - 
\mathcal{V}_{T}\left( f_{\textbf{X}^{(n)}}, f_{\tilde{\textbf{X}}^{(n)}} \right)
\\ \stackrel{(b)}{\geq}&
1 - 
\sqrt{\frac{1}{2} \dive{KL}{f_{\textbf{X}^{(n)}}}{f_{\tilde{\textbf{X}}^{(n)}}}}
\\ \stackrel{(c)}{\geq}&
1 - 
\sqrt{\frac{1}{4} \log\frac{1}{1 - a^{2}}}
\\ \stackrel{(d)}{\geq}&
1 - 
\varepsilon,
\end{align*}

where (a) is due to \cref{lemma:Connection_between_error_probabilities_to_total_variation}, (b) is due to \cref{lemma:Connection_between_divergence_to_total_variation}, (c) is by \cref{claim:Divergence_of_controlled_and_uncontrolled_AR_1_of_stable_system_in_steady_state}, and (d) if $a$ satisfies the criterion in the theorem.
\end{proof}

Next, we introduce a negative result for the setting given in \cref{theorem:Achievable_gain_using_threshold_controller}. 
\Cref{theorem:Willie_observes_the_systems_output_for_n_samples_threshold_controller} asserts that in the case of an inherently stable Gaussian AR(1) system, controlled once, i.e., reset to its stationary distribution by the threshold controller, if Willie observes the system's output via a clean channel for $n$ consecutive samples, then Willie can detect Alice's control action, for values of the gain given in the theorem.
\begin{theorem}\label{theorem:Willie_observes_the_systems_output_for_n_samples_threshold_controller}
Consider a Gaussian AR(1) system with $|a| < 1$. Alice uses the threshold controller to reset the system to its stationary distribution. Willie observes the system's output through a clean channel for $n$ samples. 
If Willie knows the one time sample in which the system is being resets, i.e., $\tau + 1$, where $X_{\tau + 1} = a \tilde{X} + Z_{\tau + 1}$, and the system's gain, then, there exists a detection method in which Willie achieves an $1-\delta$-detection for any gain satisfying
\begin{align*}
|a| 
\geq 
\sqrt{
\frac{
\left(
Q^{-1} \left( \frac{\delta}{4} \right)
\right)^{2}
-
\left(
Q^{-1}\left( \frac{2 - \delta}{4} \right)
\right)^{2}
}{
\left(
Q^{-1} \left( \frac{\delta}{4} \right)
\right)^{2}
+
\left(
Q^{-1}\left( \frac{2 - \delta}{4} \right)
\right)^{2}
}}.
\end{align*}
\end{theorem}

\begin{proof}(\cref{theorem:Willie_observes_the_systems_output_for_n_samples_threshold_controller})
Consider the following hypotheses testing problem that Willie uses in order to detect Alice's control action using the threshold controller,
\begin{align*}
\begin{matrix}
\mathcal{H}_{0}: &\textbf{X}^{(n)} & \sim & \mathcal{N}(\textbf{0}, \bs{\Sigma}_{0})
\\
\mathcal{H}_{1}: &\textbf{X}^{(n)} | \tau & \sim & \mathcal{N}(\textbf{0}, \bs{\Sigma}_{1})
\end{matrix},
\end{align*}

where $[\bs{\Sigma}_{0}]_{i,j} = \frac{\sigma_{Z}^{2}}{1 - a^{2}} a^{|i-j|}$ for $1 \leq i,j \leq n$, and 
\begin{align*}
\bs{\Sigma}_{1}
=&
\begin{pmatrix}
\bs{\Sigma}_{\tau} & \textbf{0}_{\tau, n - \tau}
\\
\textbf{0}_{n - \tau, \tau} & \bs{\Sigma}_{n-\tau}
\end{pmatrix},
\end{align*}

where $[\bs{\Sigma}_{\tau}]_{i,j} = \frac{\sigma_{Z}^{2}}{1 - a^{2}} a^{|i-j|}$ for $1 \leq i,j \leq \tau$,  and $[\bs{\Sigma}_{n - \tau}]_{i,j} = \frac{\sigma_{Z}^{2}}{1 - a^{2}} a^{|i-j|}$ for $1 \leq i,j \leq n-\tau$ (see \cref{fact:Cond_PDF_of_controlled_AR_1_process}). 

The log-likelihood ratio test is given by
\begin{align*}
T
\triangleq
\textbf{x}^{T} \left( \bs{\Sigma}_{0}^{-1} - \bs{\Sigma}_{1}^{-1} \right) \textbf{x}
\DR{\mathcal{H}_{1}}{\mathcal{H}_{0}}
2\log{t} - \log{(1 - a^{2})}
\triangleq
t',
\end{align*}

where $\log{\Big( \frac{|\bs{\Sigma}_{0}|}{|\bs{\Sigma}_{1}|} \Big)} = \log{(1 - a^{2})}$ (by
the proof of \cref{claim:Divergence_of_controlled_and_uncontrolled_AR_1_of_stable_system_in_steady_state}). 

By the proof of \cref{claim:trace_B_inv_A}, 
the inverse of $[\bs{\Sigma}_{k}]_{i,j} = \frac{\sigma_{Z}^{2}}{1 - a^{2}} a^{|i-j|}$ for $1 \leq i,j \leq k$, is given by 
\begin{align*}
[\bs{\Sigma}_{k}^{-1}]_{i,j}
=&
\frac{
\left( 1 + a^{2} \indicator{2 \leq i \leq k-1} \right) \indicator{i=j} - a \indicator{|i-j| = 1} }{\sigma_{Z}^{2}}.
\end{align*}


Hence, 
\begin{align*}
[\bs{\Sigma}_{0}^{-1} - \bs{\Sigma}_{1}^{-1}]_{i,j}
=&
\frac{a^{2}}{\sigma_{Z}^{2}}
\left( \indicator{i,j = \tau} +  \indicator{i,j = \tau + 1} \right) 
\\&- 
\frac{a}{\sigma_{Z}^{2}} \left( \indicator{i = \tau, j = \tau + 1} + \indicator{i = \tau + 1, j = \tau} \right)
.
\end{align*}


Therefore, the test, $T$, can be written explicitly as follows
\begin{align*}
T
\triangleq&
\textbf{x}^{T} \left( \bs{\Sigma}_{0}^{-1} - \bs{\Sigma}_{1}^{-1} \right) \textbf{x}
\\=&
\frac{1}{\sigma_{Z}^{2}} 
\left( 
a^{2} x_{\tau}^{2} + a^{2} x_{\tau + 1}^{2} - 2a x_{\tau} x_{\tau + 1}
\right)
\\=&
\frac{1}{\sigma_{Z}^{2}} 
\left[
\left(
x_{\tau + 1} - a x_{\tau}
\right)^{2}
-
(1 - a^{2}) x_{\tau + 1}^{2}
\right].
\end{align*}

Under each hypothesis, the test is a linear combination of two dependent chi-square random variables. 
Since under the alternative hypothesis $X_{\tau + 1} = a \tilde{X} + Z_{\tau + 1}$, where $U_{\tau + 1} = a X_{\tau} - a \tilde{X}$ and $\tilde{X} \sim \mathcal{N}\Big( 0, \frac{\sigma_{Z}^{2}}{1 - a^{2}} \Big)$.
In order to evaluate the error probabilities under each hypothesis, we need to know the distribution of the test in both cases.  
However, the distribution of $T$ is hard to evaluate under both hypotheses, hence we look at following sub-optimal test
\begin{align*}
\Bar{T}
\triangleq&
\frac{1}{\sigma_{Z}^{2}} 
\left(
x_{\tau + 1} - a x_{\tau}
\right)^{2},
\end{align*}

where $\Bar{T}$ approaches $T$ as $a \to 1^{-}$. We compare $\Bar{T}$ to a positive threshold $t^{2}$, hence, we can evaluate $\alpha$ and $\beta$ exactly. 
First, the false alarm probability is
\begin{align*}
\alpha 
=&
\prob{\Bar{T} > t^{2} | \mathcal{H}_{0}}
\stackrel{(a)}{=}
2 Q(t),
\end{align*}

where $\Bar{T} | \mathcal{H}_{0} \sim \chi_{1}^{2}$, and (a) is by \cite[equation (2.11)]{Kay_detection}.
On the other hand, the miss detection probability is 
\begin{align*}
\beta
=&
\prob{\Bar{T} \leq t^{2} | \mathcal{H}_{1}}
=
1 - 2 Q \left( \frac{t}{\sqrt{\frac{1 + a^{2}}{1 - a^{2}}}} \right),
\end{align*}

where $\frac{\Bar{T}}{\frac{1 + a^{2}}{1 - a^{2}}} | \mathcal{H}_{1} \sim \chi_{1}^{2}$.

Bounding $\alpha, \beta$ above by $\frac{\delta}{2}$, yields the requirement for $\alpha$
\begin{align}\label{eq:t_by_alpha}
t
\geq
Q^{-1}\left( \frac{\delta}{4} \right).
\end{align}

For $\beta$, we require
\begin{align}\label{eq:t_by_beta}
t
\leq
\sqrt{\frac{1 + a^{2}}{1 - a^{2}}} Q^{-1}\left( \frac{2 - \delta}{4} \right).
\end{align}

By \cref{eq:t_by_alpha,eq:t_by_beta}, Willie has to choose $t$ according to 
\begin{align}\label{eq:t'_by_beta}
Q^{-1} \left( \frac{\delta}{4} \right)
\leq
t
\leq
\sqrt{\frac{1 + a^{2}}{1 - a^{2}}} Q^{-1}\left( \frac{2 - \delta}{4} \right),
\end{align}

which yields the equivalent requirement,
\begin{align*}
|a| 
\geq 
\sqrt{
\frac{
\left(
Q^{-1} \left( \frac{\delta}{4} \right)
\right)^{2}
-
\left(
Q^{-1}\left( \frac{2 - \delta}{4} \right)
\right)^{2}
}{
\left(
Q^{-1} \left( \frac{\delta}{4} \right)
\right)^{2}
+
\left(
Q^{-1}\left( \frac{2 - \delta}{4} \right)
\right)^{2}
}}.
\end{align*}
\end{proof}

\begin{appendix}
\subsection{Proof of \cref{lemma:PDF_of_n_samples_AR_1_process}}
\label[appendix]{appendix:1}
We start with the following supporting claim.
\begin{claim}\label{claim:System_state}
The linear system shown in \cref{eq:simple_linear_model}, initializes at $n = 0$ with an initial state $X_{0}$, 
can be represented for an arbitrary controller $\{U_{n}\}$ as follows,
\begin{align}\label{eq:System_state}
X_{n} = a^{n} X_{0} + \sum_{k = 1}^{n}{a^{n - k} \left( Z_{k} -  U_{k} \right)}, \quad  \forall n \in \mathbb{N}, \ a \in \mathbb{R},
\end{align}

where $X_{n}$ is the system state at time $n$.
\end{claim}

\begin{proof}(\Cref{claim:System_state})
The proof is by induction. Clearly, for $n = 1$ the claim holds since,
\begin{align*}
X_{1}
=&
a^{1} X_{0} + 
\sum_{k = 1}^{1}{a^{1 - k} \left( Z_{k} -  U_{k} \right)}
\\=&
a^{1} X_{0} +  a^{0} (Z_{1} -  U_{1})
\\=&
a X_{0} + Z_{1} - U_{1}, 
\end{align*}

Now, for $n+1$, we have
\begin{align*}
& a^{n+1} X_{0} +
\sum_{k = 1}^{n+1}{a^{n + 1 - k} \left( Z_{k} -  U_{k} \right)}
\\=&
a^{n+1} X_{0} + 
\sum_{k = 1}^{n}{a^{n + 1 - k} \left( Z_{k} -  U_{k} \right)} 
+ Z_{n + 1} - U_{n + 1}
\\=&
a \left[ a^{n} X_{0} + \sum_{k = 1}^{n}{a^{n - k} \left( Z_{k} -  U_{k} \right)}\right] 
+ Z_{n + 1} - U_{n + 1}
\\=&
a X_{n} + Z_{n + 1} - U_{n + 1}
\\\triangleq&
X_{n + 1}.
\end{align*}

Hence, by mathematical induction \cref{eq:System_state} holds. 
\end{proof}

\begin{proof}(\Cref{lemma:PDF_of_n_samples_AR_1_process})
One can see that 
\cref{claim:System_state} can be represented with $U_{n} = 0$, as follows
\begin{align*}
\textbf{X}^{(n)}
=&
\underbrace{
\left(
\begin{matrix}
1 & 0 & 0 & \cdots & 0 \\
a & 1 & 0 & \cdots & 0 \\
a^{2} & a & 1 & \cdots & 0 \\
\vdots & \vdots & \vdots & \ddots & \vdots \\
a^{n-1} & a^{n-2} & a^{n-3} & \cdots & 1 \\
\end{matrix}
\right)}_{\textbf{A}}
\underbrace{
\left(
\begin{matrix}
Z_{1} \\
Z_{2} \\
Z_{3} \\
\vdots \\
Z_{n}
\end{matrix}
\right)}_{\textbf{Z}^{(n)}}
+
\underbrace{
\begin{pmatrix}
a \\
a^{2} \\ 
a^{3} \\
\vdots \\
a^{n}
\end{pmatrix}
}_{\tilde{\textbf{a}}^{(n)}}
X_{0}
\\=&
\textbf{A} \textbf{Z}^{(n)} + \tilde{\textbf{a}}^{(n)}
X_{0},
\end{align*}

where $\textbf{Z}^{(n)} \sim \mathcal{N}(\textbf{0}, \sigma_{Z}^{2} \textbf{I}_{n})$ and $X_{0} \sim \mathcal{N}\big( 0, \sigma_{0}^{2} \big)$, where $\textbf{I}_{n}$ is the identity matrix of size $n$. 
Hence, we have 
$\textbf{X}^{(n)} \sim \mathcal{N}(\bs{\mu}, \bs{\Sigma})$, 
when 
\begin{align*}
\mathbb{E} \left[ \textbf{X}^{(n)} \right]
=&
\textbf{A} \expc{}{}{\textbf{Z}} + \tilde{\textbf{a}}
\expc{}{}{X_{0}}
=
\textbf{0},
\\
\cov{}{}{\textbf{X}^{(n)}}
=&
\expc{}{}{ \textbf{X} \textbf{X}^{T} }
\\=&
\expc{}{}{
(\textbf{A} \textbf{Z} + \tilde{\textbf{a}}
X_{0}) 
( \textbf{Z}^{T} \textbf{A}^{T} + X_{0} \tilde{\textbf{a}}^{T} )
}
\\\stackrel{(a)}{=}&
\textbf{A} \expc{}{}{\textbf{Z} \textbf{Z}^{T}} \textbf{A}^{T}
+
\expc{}{}{X_{0}^{2}} \tilde{\textbf{a}} \tilde{\textbf{a}}^{T}
\\\stackrel{(b)}{=}&
\sigma_{Z}^{2} \textbf{A} \textbf{A}^{T} + \sigma_{0}^{2} \tilde{\textbf{a}} \tilde{\textbf{a}}^{T},
\end{align*}

where we omit the $(\cdot)^{(n)}$ notation for simplicity, (a) is since $X_{0}$ is independent of $\textbf{Z}$ and both with expectation of zero, and (b) is due to 
$\expc{}{}{X_{0}^{2}} = \sigma_{0}^{2}$. 
In addition, take $\textbf{y} \in \mathbb{R}^{n} \backslash \big\{ \textbf{0}^{(n)} \big\}$, hence
\begin{align*}
\textbf{y}^{T} \bs{\Sigma} \textbf{y}
=&
\textbf{y}^{T} (\sigma_{Z}^{2} \textbf{A} \textbf{A}^{T} + \sigma_{0}^{2} \tilde{\textbf{a}} \tilde{\textbf{a}}^{T}) \textbf{y}
\\=&
\sigma_{Z}^{2} (\textbf{A}^{T} \textbf{y})^{T} (\textbf{A}^{T} \textbf{y}) + \sigma_{0}^{2} \textbf{y}^{T} \tilde{\textbf{a}} (\textbf{y}^{T} \tilde{\textbf{a}})^{T} \\=&
\sigma_{Z}^{2} \Vert \textbf{A}^{T} \textbf{y} \Vert^{2} + \sigma_{0}^{2} (\textbf{y}^{T} \tilde{\textbf{a}})^{2}
\stackrel{(a)}{>}
0,
\end{align*}

where (a) is by a property of the rank of a matrix,
$\operatorname*{rank}(\textbf{A} \textbf{A}^{T}) = \operatorname*{rank}(\textbf{A}) = n$. 
Therefore, $\bs{\Sigma}$ 
is invertible, and \cref{eq:PDF_of_n_samples_AR_1_process} 
is well-defined. 

Furthermore, $[\textbf{A}]_{i,j} = a^{i-j} u(i-j)$ and $[\tilde{\textbf{a}} \tilde{\textbf{a}}^{T}]_{i,j} = a^{i+j}$, where $u(\cdot)$ is the discrete step function. Hence, we have
\begin{align*}
[\bs{\Sigma}]_{i,j}
=&
\sigma_{Z}^{2} [\textbf{A} \textbf{A}^{T}]_{i,j} + \sigma_{0}^{2} [\tilde{\textbf{a}} \tilde{\textbf{a}}^{T}]_{i,j}
\\=&
\sigma_{Z}^{2} \sum_{k=1}^{n}{[\textbf{A}]_{i,k} [\textbf{A}]_{j,k}}
+
\sigma_{0}^{2} a^{i+j}
\\=&
\sigma_{Z}^{2} \sum_{k=1}^{n}{a^{i-k} u(i-k) a^{j-k} u(j-k)}
+
\sigma_{0}^{2} a^{i+j}
\\\stackrel{(a)}{=}&
\sigma_{Z}^{2} a^{i+j} \sum_{k=1}^{\min(i,j)}{a^{-2k}}
+
\sigma_{0}^{2} a^{i+j}
\\=&
\sigma_{Z}^{2} a^{i+j} \frac{a^{-2 \min(i,j)} - 1}{1 - a^{2}}
+
\sigma_{0}^{2} a^{i+j}
\\\stackrel{(b)}{=}&
\frac{\sigma_{Z}^{2}}{1 - a^{2}} \left( a^{|i-j|} - a^{i+j} \right) + \sigma_{0}^{2} a^{i+j}, 
\quad  1 \leq i,j \leq n,
\end{align*}

where (a) is since $i \geq k$ and $j \geq k$, hence, 
$k \leq \min(i,j)$. (b) is since $i+j-2 \min(i,j) = |i-j|$.

For the special case in which $|a| < 1$, $\sigma_{0}^{2} = \frac{\sigma_{Z}^{2}}{1 - a^{2}}$ and $X_{0} \sim \mathcal{N}\Big( 0, \frac{\sigma_{Z}^{2}}{1 - a^{2}} \Big)$, i.e., $X_{0}$ drawn according to the stationary distribution, we have 
\begin{align*}
[\bs{\Sigma}]_{i,j}
=&
\frac{\sigma_{Z}^{2}}{1 - a^{2}} a^{|i-j|}, 
\quad 1 \leq i,j \leq n.
\end{align*}
\end{proof}
\subsection{Proof of \Cref{lemma:Divergence_of_gaussian_vector}}
\label[appendix]{appendix:2}
\begin{proof}(\Cref{lemma:Divergence_of_gaussian_vector})
It is easy to see that, 
\begin{align*}
\log \left(
\frac{\mathcal{N}_{0}}{\mathcal{N}_{1}}
\right)
=&
\frac{1}{2} \Bigg(
\log \frac{|\bs{\Sigma}_{1}|}{|\bs{\Sigma}_{0}|} 
+
(\textbf{x} - \bs{\mu}_{1})^{T} \bs{\Sigma}_{1}^{-1} (\textbf{x} - \bs{\mu}_{1}) 
\\&- 
(\textbf{x} - \bs{\mu}_{0})^{T} \bs{\Sigma}_{0}^{-1} (\textbf{x} - \bs{\mu}_{0}) \Bigg). 
\end{align*}

Applying expectation with respect to $\mathcal{N}_{0}$ yields, 
\begin{align*}
\expc{\mathcal{N}_{0}}{}{\log \left(
\frac{\mathcal{N}_{0}}{\mathcal{N}_{1}}
\right)}
=&
\frac{1}{2} \expc{\mathcal{N}_{0}}{}{(\textbf{x} - \bs{\mu}_{1})^{T} \bs{\Sigma}_{1}^{-1} (\textbf{x} - \bs{\mu}_{1})}
\\&- 
\frac{1}{2} \expc{\mathcal{N}_{0}}{}{(\textbf{x} - \bs{\mu}_{0})^{T} \bs{\Sigma}_{0}^{-1} (\textbf{x} - \bs{\mu}_{0})}
\\&+
\frac{1}{2}
\log \frac{|\bs{\Sigma}_{1}|}{|\bs{\Sigma}_{0}|}
\end{align*}

Solving each of the expectations above, 
\begin{align*}
\mathbb{E}_{\mathcal{N}_{0}} & \left[ 
(\textbf{x} - \bs{\mu}_{0})^{T} \bs{\Sigma}_{0}^{-1} (\textbf{x} - \bs{\mu}_{0})
\right]
\\=&
\expc{\mathcal{N}_{0}}{}{\tr{(\textbf{x} - \bs{\mu}_{0})^{T} \bs{\Sigma}_{0}^{-1} (\textbf{x} - \bs{\mu}_{0})}}
\\=&
\expc{\mathcal{N}_{0}}{}{\tr{\bs{\Sigma}_{0}^{-1} (\textbf{x} - \bs{\mu}_{0})(\textbf{x} - \bs{\mu}_{0})^{T}}}
\\=&
\tr{\bs{\Sigma}_{0}^{-1} \expc{\mathcal{N}_{0}}{}{ (\textbf{x} - \bs{\mu}_{0})(\textbf{x} - \bs{\mu}_{0})^{T} }}
\\=&
\tr{\bs{\Sigma}_{0}^{-1} \bs{\Sigma}_{0}}
\\=&
n, 
\end{align*}

second, 
\begin{align*}
\mathbb{E}_{\mathcal{N}_{0}} & \left[ 
(\textbf{x} - \bs{\mu}_{1})^{T} \bs{\Sigma}_{1}^{-1} (\textbf{x} - \bs{\mu}_{1})
\right]
\\=&
\tr{\bs{\Sigma}_{1}^{-1} \expc{\mathcal{N}_{0}}{}{ (\textbf{x} - \bs{\mu}_{1})(\textbf{x} - \bs{\mu}_{1})^{T}}}
\\\stackrel{(a)}{=}&
\tr{\bs{\Sigma}_{1}^{-1} (\bs{\Sigma_{0}} + (\bs{\mu}_{1} - \bs{\mu}_{0}) (\bs{\mu}_{1} - \bs{\mu}_{0})^{T})}
\\=&
\tr{\bs{\Sigma}_{1}^{-1} \bs{\Sigma_{0}}} + (\bs{\mu}_{1} - \bs{\mu}_{0})^{T} \bs{\Sigma}_{1}^{-1} (\bs{\mu}_{1} - \bs{\mu}_{0}), 
\end{align*}

where (a) is since, 
\begin{align*}
\mathbb{E}_{\mathcal{N}_{0}} & \left[ 
(\textbf{x} - \bs{\mu}_{1})(\textbf{x} - \bs{\mu}_{1})^{T}
\right]
\\=&
\expc{\mathcal{N}_{0}}{}{(\textbf{x} - \bs{\mu}_{0} +\bs{\mu}_{0} - \bs{\mu}_{1})  (\textbf{x} - \bs{\mu}_{0} +\bs{\mu}_{0} - \bs{\mu}_{1})^{T}}
\\=&
\expc{\mathcal{N}_{0}}{}{(\textbf{x} - \bs{\mu}_{0}) (\textbf{x} - \bs{\mu}_{0})^{T}}
+
\expc{\mathcal{N}_{0}}{}{(\textbf{x} - \bs{\mu}_{0})} (\bs{\mu}_{0} - \bs{\mu}_{1})^{T} 
\\&+
(\bs{\mu}_{0} - \bs{\mu}_{1}) \expc{\mathcal{N}_{0}}{}{(\textbf{x} - \bs{\mu}_{0})^{T}} 
+
(\bs{\mu}_{0} - \bs{\mu}_{1}) (\bs{\mu}_{0} - \bs{\mu}_{1})^{T}
\\=&
\bs{\Sigma_{0}} + (\bs{\mu}_{1} - \bs{\mu}_{0}) (\bs{\mu}_{1} - \bs{\mu}_{0})^{T}.
\end{align*}
\end{proof}

\subsection{Proof of \cref{claim:One_bit_controller_representation}}
\label[appendix]{appendix:3}
\begin{claim}\label{claim:One_bit_controller_representation}
The \emph{One-bit controller}, shown in \cref{eq:One_bit_controller}, can be written as,
\begin{align*}
U_{n}
=
\left[
\frac{a}{2} \frac{B}{1 - a/2} + \left(\frac{a}{2}\right)^{n-1} \left( C_{1} - \frac{B}{1 - a/2}\right)
\right] \operatorname*{sgn}(X_{n-1}),
\end{align*}

where 
$C_{1} \geq \frac{B}{1 - a/2}$ 
is the first element in the series $C_{n}$, 
$B$ is the bound of the noise $\{Z_{n}\}$, .i.e., $|Z_{n}| \leq B, \ \forall n \geq 1$ and  
$a \neq 2$ is the gain of the system shown in  \cref{eq:simple_linear_model}.
\end{claim}

\begin{proof}(\Cref{claim:One_bit_controller_representation})
By \cref{eq:One_bit_controller}, one has to show that the following holds,
\begin{align*}
C_{n-1}
=
\frac{B}{1 - a/2} + \left(\frac{a}{2}\right)^{n-2} \left( C_{1} - \frac{B}{1 - a/2}\right).
\end{align*}

First, we will prove the following by induction,
\begin{align}\label{eq:One_bit_controller_C_n-k}
C_{n}
=
\frac{B}{1 - a/2}
+ \left(\frac{a}{2}\right)^{k} \left( C_{n-k} - \frac{B}{1 - a/2} \right),
\end{align}

where $k$ is some shift of the series $C_{n}$. For instance, if $k=1$ we have $C_{n} = (a/2) C_{n-1} + B$, which is by definition. Assume \cref{eq:One_bit_controller_C_n-k} holds for 
$k=r$, then for $k= r+1$,
\begin{align*}
C_{n}
=&
\frac{B}{1 - a/2}
+ \left(\frac{a}{2}\right)^{r} \left( C_{n-r} - \frac{B}{1 - a/2} \right)
\\\stackrel{(a)}{=}&
\frac{B}{1 - a/2}
+ \left(\frac{a}{2}\right)^{r} \left( \left( \frac{a}{2} C_{n-r-1} + B \right) - \frac{B}{1 - a/2} \right)
\\=&
\frac{B}{1 - a/2}
+ \left(\frac{a}{2}\right)^{r} \left( \frac{a}{2} C_{n-r-1} - \frac{a}{2} \frac{B}{1 - a/2} \right)
\\=&
\frac{B}{1 - a/2}
+ \left(\frac{a}{2}\right)^{r+1} \left( C_{n-(r+1)} - \frac{B}{1 - a/2} \right),
\end{align*}

where (a) is by the relation: $C_{n} = (a/2) C_{n-1} + B$. 
Therefore, \cref{eq:One_bit_controller_C_n-k} holds $\forall 1 \leq k < n$. 
Let us substitute to \cref{eq:One_bit_controller_C_n-k}, $k = n-1$,
\begin{align*}
\begin{split}
C_{n}
=&
\frac{B}{1 - a/2}
+ \left(\frac{a}{2}\right)^{n-1} \left( C_{n-(n-1)} - \frac{B}{1 - a/2} \right)
\\=&
\frac{B}{1 - a/2}
+ \left(\frac{a}{2}\right)^{n-1} \left( C_{1} - \frac{B}{1 - a/2} \right).
\end{split}
\end{align*}
\end{proof}

\begin{remarkc}
By setting $C_{1} = \frac{B}{1 - a/2}$,
\begin{align}\label{eq:One_bit_controller_representation_spcial_case}
U_{n}
=
\frac{a}{2} \frac{B}{1 - a/2} \operatorname*{sgn}(X_{n-1}), \quad \forall n > 1,
\end{align}

since $C_{n}$ is monotonically decreasing to $\frac{B}{1 - a/2}$, hence,
\begin{align}\label{eq:One_bit_controller_representation_spcial_case_abs_value}
|U_{n}|
=
\frac{a}{2} \frac{B}{1 - a/2}, \quad \forall n > 1,
\end{align}

since $a/2 < 1$. 
As $n \to \infty$, \crefrange{eq:One_bit_controller_representation_spcial_case}{eq:One_bit_controller_representation_spcial_case_abs_value} also holds regardless of the choice of $C_{1}$ (however, it has to be greater or equal to $\frac{B}{1 - a/2}$). 
Thus, one can deduce that $|U_{n}|$ 
converges. I.e., $|U_{n}|$ can not get arbitrary large.
\end{remarkc}

Proceeding to the proof of \cref{claim:One_bit_controller_energy}.

\begin{proof}(\Cref{claim:One_bit_controller_energy})
we have,
\begin{align*}
E_{U} 
\triangleq&
\frac{1}{N}\sum_{n = 1}^{N}{U_{n}^{2}}
\\=&
\frac{1}{N}\sum_{n = 1}^{N}{\frac{a^{2}}{4} C_{n-1}^{2} (\operatorname*{sgn}(X_{n-1}))^{2}}
\\=&
\frac{a^{2}}{4} \frac{1}{N}\sum_{n = 1}^{N}{C_{n-1}^{2}}
\\\stackrel{(a)}{\leq}&
\frac{a^{2}}{4} \frac{1}{N}\sum_{n = 1}^{N}{C_{1}^{2}}
\\=&
\left( \frac{a}{2} C_{1} \right)^{2},
\end{align*}

where (a) is since $C_{n}$ is a monotonically decreasing series. 

On the other hand,
\begin{align*}
E_{U} 
=&
\frac{a^{2}}{4} \frac{1}{N}\sum_{n = 1}^{N}{C_{n-1}^{2}}
\\\stackrel{(a)}{\geq}&
\frac{a^{2}}{4} \frac{1}{N}\sum_{n = 1}^{N}{\left( \frac{B}{1 - a/2} \right)^{2}}
\\=&
\frac{a^{2}}{4} \left( \frac{B}{1 - a/2} \right)^{2}
\\=&
\left( \frac{a B}{2 - a} \right)^{2},
\end{align*}

where (a) is since $C_{n}$ is a monotonically decreasing series converging to: $\frac{B}{1 - a/2}$.
\end{proof}
 

\subsection{Proof of \Cref{claim:trace_B_inv_A}}
\label[appendix]{appendix:4}
For $|a|, |b| < 1$, and in steady state, by  \cref{lemma:PDF_of_n_samples_AR_1_process}, 
$[\bs{\Sigma_{0}}]_{i,j} = \frac{\sigma_{Z}^{2}}{1 - a^{2}} a^{|i-j|}$ and 
$[\bs{\Sigma_{1}}]_{i,j} = \frac{\sigma_{Z}^{2}}{1 - b^{2}} b^{|i-j|}$ for $1 \leq i,j \leq n$. 
Denote $[\textbf{A}]_{i,j} = a^{|i-j|}$ and 
$[\textbf{B}]_{i,j} = b^{|i-j|}$ for $1 \leq i,j \leq n$. 
To show that 
\begin{align}\label{eq:A_inv}
[\textbf{A}^{-1}]_{i,j}
=&
\frac{
\left( 1 + a^{2} \indicator{2 \leq i \leq n-1} \right) \indicator{i=j} - a \indicator{|i - j| = 1} 
}{1 - a^{2}},
\end{align}


we check by definition that $\textbf{A}^{-1} \textbf{A} = \textbf{I}_{n}$. 
First, 
\begin{align*}
[\textbf{A}^{-1} \textbf{A}]_{i,j}
=&
\sum_{k=1}^{n}{[\textbf{A}]_{i,k} [\textbf{A}^{-1}]_{k,j}}
\\=&
\frac{1}{1 - a^{2}}
\sum_{k=1}^{n}{\left[
\left( 1 + a^{2} \indicator{2 \leq k \leq n-1} \right) \indicator{k=j} a^{|i-k|} 
\right.}
\\& \left.
- a \left( \indicator{k = j-1} + \indicator{k = j+1} \right) a^{|i-k|} \right]
\\=&
\frac{1}{1 - a^{2}}
\left[
\left( 1 + a^{2} \indicator{2 \leq j \leq n-1} \right) a^{|i-j|} 
\right.
\\& \left.
- a \left( a^{|i-j + 1|} \indicator{j \neq 1} + a^{|i-j - 1|} \indicator{j \neq n} \right) \right],
\end{align*}

for $i \geq j + 1$, $2 \leq i \leq n$ and $1 \leq j \leq n-1$, thus
\begin{align*}
[\textbf{A}^{-1} \textbf{A}]_{i,j}
=&
\frac{\indicator{1 \leq j \leq n-1} \indicator{2 \leq i \leq n}}{1 - a^{2}}
\left[
\left( 1 + a^{2} \indicator{2 \leq j \leq n-1} \right) a^{i-j} 
\right.
\\& \left.
- a \left( a^{i-j + 1} \indicator{j \neq 1} + a^{i-j - 1} \indicator{j \neq n} \right) \right]
\\=&
\frac{\indicator{1 \leq j \leq n-1} \indicator{2 \leq i \leq n}}{1 - a^{2}}
\left[
a^{i-j} \indicator{j = n} 
\right.
\\& \left. -
a^{i-j+2} \indicator{j = n} 
\right]
= 0,
\end{align*}

Similarly for $j \geq i +1$, we have $[\textbf{A}^{-1} \textbf{A}]_{i,j} = 0$. For $i = j$, 
\begin{align*}
[\textbf{A}^{-1} \textbf{A}]_{i,i}
=&
\frac{1}{1 - a^{2}}
\left[
\left( 1 + a^{2} \indicator{2 \leq j \leq n-1} \right) 
\right.
\\& \left.
- a \left( a \indicator{j \neq 1} + a \indicator{j \neq n} \right) \right]
\\=&
\frac{1}{1 - a^{2}}
\left[
1 - a^{2}
\right]
= 1.
\end{align*}

Now, by changing $a$ to $b$ in $[\textbf{A}^{-1}]_{i,j}$,  $[\textbf{B}^{-1}]_{i,j}$ is obtained. 
Next, we evaluate the following trace 
\begin{align*}
\tr{\textbf{B}^{-1} \textbf{A}}
=&
\sum_{m=1}^{n}{
\sum_{k=1}^{n}{
[\textbf{B}^{-1}]_{m,k} [\textbf{A}]_{k,m}}}
\\=&
\frac{1}{1 - b^{2}}
\sum_{m=1}^{n}{
\sum_{k=1}^{n}{
\left[ 
-b a^{|k-m|} \indicator{|k-m|=1}
\right.}}
\\& 
+ b^{2} a^{|k-m|} \indicator{2 \leq m \leq n-1} \indicator{k=m}
\\& \left.
+ a^{|k-m|} \indicator{k=m}
\right]
\\=&
\frac{1}{1 - b^{2}}
\sum_{m=1}^{n}{
\left[ 
-ab (\indicator{m \neq 1} + \indicator{m \neq n})
\right.}
\\&+ \left.
b^{2} \indicator{2 \leq m \leq n-1}
+ 1
\right]
\\=&
\frac{(n-2)b^{2} - 2(n-1) ab + n}{1 - b^{2}}.
\end{align*}

Thus, 
\begin{align*}
\tr{\bs{\Sigma}_{1}^{-1} \bs{\Sigma}_{0}}
=&
\frac{1 - b^{2}}{1 - a^{2}}
\tr{\textbf{B}^{-1} \textbf{A}}
\\=&
\frac{(n-2) b^{2} - 2(n-1) ab + n}{1 - a^{2}}.
\end{align*}

\subsection{Proof of  \Cref{claim:Divergence_of_controlled_and_uncontrolled_AR_1_of_stable_system_in_steady_state}}
\label[appendix]{appendix:5}
By \cref{lemma:Divergence_of_gaussian_vector}, the bound in \cref{eq:Divergence_of_controlled_and_uncontrolled_AR_1} is equivalent to 
\begin{align*}
\frac{1}{2}
\expc{p_{\tau}}{}{
\tr{
\tilde{\bs{\Sigma}}_{n}^{-1} \bs{\Sigma}_{n}} 
+ \log{
\frac{\left\vert \tilde{\bs{\Sigma}}_{n} \right\vert}
{\left\vert \bs{\Sigma}_{n} \right\vert}}} -\frac{n}{2}.
\end{align*}

Thus, we need to show two things. First
\begin{align}\label{eq:trace_of_invrese_C_1_C_0}
\tr{\tilde{\bs{\Sigma}}_{n}^{-1} \bs{\Sigma}_{n}} = n,
\end{align}
and second, 
\begin{align}\label{eq:log_det_S_1_over_det_S0}
\log{\frac{\left\vert \tilde{\bs{\Sigma}}_{n}  \right\vert}
{\left\vert \bs{\Sigma}_{n} \right\vert}} 
=
\log{\frac{1}{1 - a^{2}}}.
\end{align}

Recall that an $n$-tuple of a Gaussian AR(1) process with $|a| < 1$ at steady state, has the following covariance matrix
$[\bs{\Sigma}]_{i,j} = \frac{\sigma_{Z}^{2}}{1 - a^{2}} a^{|i-j|}$. 
Hence, in our case, $\left[  \bs{\Sigma}_{n} \right]_{i,j} = \frac{\sigma_{Z}^{2}}{1 - a^{2}} a^{|i-j|}$, $1 \leq i,j \leq n$, and, $\tilde{\bs{\Sigma}}_{n} =
\begin{pmatrix}
\bs{\Sigma}_{\tau} & \textbf{0}_{\tau, n - \tau} \\
\textbf{0}_{n - \tau, \tau} & \bs{\Sigma}_{n - \tau},
\end{pmatrix}$, where,
\begin{align*}
\left[ \bs{\Sigma}_{\tau} \right]_{i,j}
=&
\frac{\sigma_{Z}^{2}}{1 - a^{2}} a^{|i-j|}, \quad 1 \leq i,j \leq \tau,
\\
\left[ \bs{\Sigma}_{n-\tau} \right]_{i,j}
=&
\frac{\sigma_{Z}^{2}}{1 - a^{2}} a^{|i-j|}, \quad 1 \leq i,j \leq n-\tau.
\end{align*}

One can divide 
$\bs{\Sigma}_{n}$ 
to blocks the same way as 
$\tilde{\bs{\Sigma}}_{n} $, 
thus, 
\begin{align*}
\bs{\Sigma}_{n}
=
\begin{pmatrix}
\bs{\Sigma}_{11} & \bs{\Sigma}_{12} \\
\bs{\Sigma}_{12}^{T} & \bs{\Sigma}_{22}
\end{pmatrix},
\end{align*}

where, 
$\left[ \bs{\Sigma}_{11} \right]_{i,j}
=
\left[ \bs{\Sigma}_{\tau} \right]_{i,j}$, and 
$\left[ \bs{\Sigma}_{22} \right]_{i,j}
=
\left[ \bs{\Sigma}_{n-\tau} \right]_{i,j}$.

Therefore,
\begin{align*}
\tilde{\bs{\Sigma}}_{n}^{-1} \bs{\Sigma}_{n}
=&
\begin{pmatrix}
\bs{\Sigma}_{\tau}^{-1} & \textbf{0}_{\tau, n - \tau} \\
\textbf{0}_{n - \tau, \tau} & \bs{\Sigma}_{n - \tau}^{-1}
\end{pmatrix}
\begin{pmatrix}
\bs{\Sigma}_{11} & \bs{\Sigma}_{12} \\
\bs{\Sigma}_{12}^{T} & \bs{\Sigma}_{22}
\end{pmatrix}
\\=&
\begin{pmatrix}
\textbf{I}_{\tau} & \bs{\Sigma}_{\tau}^{-1} \bs{\Sigma}_{12} \\
\bs{\Sigma}_{n-\tau}^{-1} \bs{\Sigma}_{12}^{T} & \textbf{I}_{n-\tau} 
\end{pmatrix},
\end{align*}

which yields \cref{eq:trace_of_invrese_C_1_C_0}.

Now, we move to the second part of the proof. 
We use a LU decomposition for the matrix defined as
$\left[ \hat{\textbf{A}}_{n} \right]_{i,j} = a^{|i-j|}, \quad 1 \leq i,j \leq n$. 
Where the lower triangular matrix $\textbf{L}_{n}$ defined as 
$\left[ \textbf{L}_{n} \right]_{i,j} = a^{i-j} u(i-j), \quad 1 \leq i,j \leq n$, 
and the upper triangular matrix $\textbf{U}_{n}$ defined as 
$\left[ \textbf{U}_{n} \right]_{i,j} = a^{j-i} (1 - a^{2})^{(1 - \delta(i-1))} u(j-i), \quad 1 \leq i,j \leq n$. 
If so, 
\begin{align*}
\left[ \hat{\textbf{A}}_{n} \right]_{i,j}
=&
\sum_{k=1}^{n}{a^{i-k} u(i-k) a^{j-k} (1 - a^{2})^{(1 - \delta(k-1))} u(j-k)}
\\\stackrel{(a)}{=}&
a^{i+j}
\sum_{k=1}^{\gamma}{a^{-2k} (1 - a^{2})^{(1 - \delta(k-1))}}
\\\stackrel{(b)}{=}&
a^{i+j}
\left(  
a^{-2} + 
(1 - a^{2}) \sum_{k=2}^{\gamma}{a^{-2k}}
\right)
\\\stackrel{(c)}{=}&
a^{i+j}
\left(  
a^{-2} + 
(1 - a^{2}) \frac{a^{-2 \gamma} - a^{-2}}{1 - a^{2}}
\right)
\\=&
a^{i+j}
\left(  
a^{-2} + a^{-2 \gamma} - a^{-2}
\right)
\\\stackrel{(d)}{=}&
a^{|i-j|},
\end{align*}

where (a) is by the fact that $k \leq i$ and $k \leq j$, hence $k \leq \min{(i,j)} \triangleq \gamma$. 
(b) is by dividing the sum for $k = 1$ and for $k > 1$. 
(c) is due to sum of geometric series. 
(d) is by $i + j - 2\min{(i,j)} = |i-j|$. 

Now, consider that $\left[ \textbf{L}_{n} \right]_{i,i} = 1$ and 
$\left[ \textbf{U}_{n} \right]_{i,i} = \delta(i-1) + (1 - a^{2}) u(i-2)$, 
which yields, 
\begin{align*}
|\hat{\textbf{A}}_{n}|
=
|\textbf{L}_{n}| |\textbf{U}_{n}|
=
1 \cdot 1 \cdot (1 - a^{2})^{n-1}
= 
(1 - a^{2})^{n-1},
\end{align*}

therefore, for $1 \leq \tau < n$ we have, 
\begin{align*}
& \log
\frac{\left\vert \tilde{\bs{\Sigma}}_{n}  \right\vert}
{\left\vert \bs{\Sigma}_{n} \right\vert} 
=
\log
\frac{\left\vert \frac{\sigma_{Z}^{2}}{1 - a^{2}} \hat{\textbf{A}}_{\tau} \right\vert \left\vert \frac{\sigma_{Z}^{2}}{1 - a^{2}} \hat{\textbf{A}}_{n - \tau} \right\vert}
{\left\vert \frac{\sigma_{Z}^{2}}{1 - a^{2}} \hat{\textbf{A}}_{n} \right\vert} 
\\=&
\log
\frac{\left( \frac{\sigma_{Z}^{2}}{1 - a^{2}} \right)^{\tau}
(1 - a^{2})^{\tau-1}
\left( \frac{\sigma_{Z}^{2}}{1 - a^{2}} \right)^{n - \tau}
(1 - a^{2})^{n - \tau-1}}
{\left( \frac{\sigma_{Z}^{2}}{1 - a^{2}} \right)^{n}
(1 - a^{2})^{n-1}} 
\\=&
\log
\frac{1}{1 - a^{2}}.
\end{align*}

\end{appendix}

\addcontentsline{toc}{section}{\numberline{}References}
\bibliographystyle{IEEEtran}
\bibliography{ON_AUTOMATIC_CONTROL/Bibliography/bibliography}
\begin{IEEEbiography}[{\includegraphics[width=1in,height=1.25in,clip,keepaspectratio]{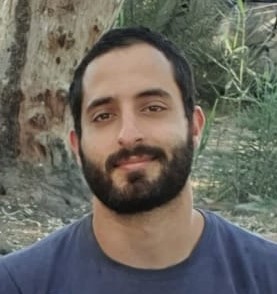}}]{Barak Amihood} 
received the B.Sc. (Summa Cum Laude) degree from the Department of Electrical and Electronics Engineering, SCE - Shamoon College of Engineering, Israel, in 2018. 
He is pursuing an M.Sc. degree from the School of Electrical and Computer Engineering, BGU - Ben Gurion University of the Negev, Israel.
\end{IEEEbiography}
\begin{IEEEbiography}[{\includegraphics[width=1in,height=1.25in,clip,keepaspectratio]{ON_AUTOMATIC_CONTROL/figures/Asaf_Cohen.jpg}}]{Asaf Cohen}
received the B.Sc. (Hons.), M.Sc. (Hons.), and Ph.D. degrees from the Department of Electrical Engineering, Technion, Israel Institute of Technology, in 2001, 2003, and 2007, respectively.
From 1998 to 2000, he was with the IBM Research Laboratory, Haifa, where he was working on distributed computing. Between 2007 and 2009 he was a Post-Doctoral Scholar at the California Institute of Technology, and between 2015–2016 he was a visiting scientist at the Massachusetts Institute of Technology. He is currently an Associate Professor at the School of Electrical and Computer Engineering, Ben-Gurion University of the Negev, Israel. His areas of interest are information theory, learning, and coding. In particular, he is interested in network information theory, network coding and coding in general, network security and anomaly detection, statistical signal processing with applications to detection and estimation and sequential decision-making.
He received several honors and awards, including the Viterbi Post-Doctoral Scholarship, the Dr. Philip Marlin Prize for Computer Engineering in 2000, the Student Paper Award from IEEE Israel in 2006 and the Ben-Gurion University Excellence in Teaching award in 2014. He served as a Technical Program Committeee for ISIT, ITW and VTC for several years, and is currently an Associate Editor for Network Information Theory and Network Coding; Physical Layer Security; Source/Channel Coding and Cross-Layer Design to the IEEE Transactions on Communications.
\end{IEEEbiography}
\end{document}